\documentclass[aps,pra,reprint,showpacs,superscriptaddress,longbibliography]{revtex4-1}
\usepackage{mathtools,amssymb,graphicx,units}
\usepackage[plainpages=false,pdfpagelabels,colorlinks=true,linkcolor=red,urlcolor=blue,citecolor=blue,pdftitle={Title},pdfauthor={},pdfdisplaydoctitle=true,pdfduplex=DuplexFlipLongEdge]{hyperref}
\usepackage{soul} % to use \st for strikeout 
\usepackage[dvipsnames,usenames]{color}
\usepackage[normalem]{ulem}
\usepackage{graphicx}
\usepackage{xcolor}
\usepackage{bbold} % to generate identity 1
\usepackage{float}
\emergencystretch=\maxdimen
\newcommand{\al}{\alpha}

\newcommand{\s}{\sigma}

\def\ave#1{\langle #1\rangle}

\newcommand{\phdag}{{\phantom{\dagger}}}

\newcommand{\iv}{\mathbf{i}}
\newcommand{\jv}{\mathbf{j}}
\newcommand{\kv}{\mathbf{k}}
\newcommand{\mv}{\mathbf{m}}
\newcommand{\qv}{\mathbf{q}}
\newcommand{\Qv}{\mathbf{Q}}
\newcommand{\Rv}{\mathbf{R}}
\newcommand{\Sv}{\mathbf{S}}

\begin{document}

\title{Magnetism and metal-insulator transitions in the Rashba-Hubbard model}
\author{Welberth \surname{Kennedy}} 
\email{welberthkennedy@icloud.com}
\affiliation{Departamento de F\'\i sica, Universidade Federal do Piau\'\i, Teresina, PI, Brazil}
\author{Sebasti\~ao dos Anjos \surname{Sousa-J\'unior}} 
\affiliation{Instituto de F\'\i sica, Universidade Federal do Rio de Janeiro,
21941-972 Rio de Janeiro RJ, Brazil}
\author{Natanael C. Costa}
\affiliation{Instituto de F\'\i sica, Universidade Federal do Rio de Janeiro,
21941-972 Rio de Janeiro RJ, Brazil}
\author{Raimundo R. \surname{dos Santos}} 
\affiliation{Instituto de F\'\i sica, Universidade Federal do Rio de Janeiro,
21941-972 Rio de Janeiro RJ, Brazil}

\begin{abstract}
The nature of metal-insulator and magnetic transitions is still a subject under intense debate in condensed matter physics. Amongst the many possible mechanisms, the interplay between electronic correlations and spin-orbit couplings is an issue of a great deal of interest, in particular when dealing with quasi-2D compounds. In view of this, here we use a Hartree-Fock approach to investigate how the Rashba spin-orbit coupling, $V_\text{SO}$, affects the magnetic ordering provided by a Hubbard interaction, $U$, on a square lattice. At half-filling, we have found a sequence of transitions for increasing $V_\text{SO}$: from a Mott insulator to a metallic antiferromagnet, and then to a paramagnetic Rashba metal. Also, our results indicate that the Rashba coupling favors magnetic striped phases in the doped regime. By analyzing spectral properties, we associate the rearrangement of the magnetic ordering with the emerging chirality created by the spin-orbit coupling.
Our findings provide insights towards clarifying the competition between these tendencies.

\end{abstract}

\maketitle

%%%%%%%%%%%%%% Section I: Introduction %%%%%%%%%%%%%%%%%%%%%%%%%%

\section{Introduction}
\label{sec:intro}

The interplay between strong electron correlations and spin-orbit coupling (SOC) has attracted a great deal of attention over recent years \cite{Witczak14,Bercioux15,Manchon15,Bertinshaw19}. 
Theoretical interest has found echo in experimental studies both in materials, such as semiconductor heterostructures, carbon-based materials, pyrochlore iridates \cite{Bertinshaw19}, and superconducting cuprates~\cite{Gotlieb18}, as well as in ultracold fermionic atoms in optical lattices~\cite{Kolkowitz17,Song19}.
One of the reasons for such activity stems from the fact that the paradigmatic Mott insulating phase is affected by the presence of a SOC in a variety of ways, giving rise to many exotic quantum states of matter, such as topological insulators, Weyl semimetals, Kitaev spin liquids, and so forth~\cite{Witczak14,Bercioux15,Manchon15,Bertinshaw19}. 
SOC has also been predicted to modify more conventional ordered phases such as magnetism and superconductivity, as well as enhancing the possibility of superconducting triplet pairing \cite{Greco18}.

Despite the intense activity, there is still no consensus on the mechanisms through which increasing SOC suppresses antiferromagnetic order (or any spiral ordering) on the way to the above mentioned more exotic phases. 
Indeed, even in the simplest case, namely that of the single-band Hubbard model, the effects of  Rashba SOC \cite{Rashba60} on the ground state phase diagram, $U\times n$ (where $U$ is the on-site repulsion, and $n$ is the band filling), are crucially dependent on the theoretical approach. 
At half filling, a cluster dynamical mean-field theory (CDMFT) predicts the existence of a metallic phase which eventually becomes an insulating phase as $U$ increases, with magnetic arrangements changing with the strength of the SOC \cite{Zhang15}.
Also at half filling, a cluster perturbation theory (CPT) recently found that the SOC favors the formation of a metallic state \cite{Brosco20}.

In view of this, we feel that a thorough examination of the ground state phase diagram of the Hubbard model with Rashba SOC is still lacking, especially in the doped case. 
With this in mind, here we use a Hartree-Fock~(HF) approach capable of detecting a diversity of spiral magnetic structures, while also shedding light on the metallic or insulating character of the phases involved.  
The layout of the paper is as follows. 
In Sec.\,\ref{sec:model} we present both the model and highlights of the HF approach, whose results are presented and discussed in Sec.\,\ref{sec:results}. 
Conclusions are presented in Sec.\,\ref{sec:concl}.
 
\color{black}
%%%%%%%%%%%%%%%%%%%%%%%%%%%%%%%%%%%%%%%%%%%%%%
\section{Model and methods}
\label{sec:model}

%%%%%%%%%%%%%%%%%%%%%%%%%%%%%%%%%%%%%%%%%%%%%
The system is described by the Hamiltonian, 
\begin{align}
\mathcal{H} = \mathcal{H}_{H} + \mathcal{H}_\text{SO},
\label{eq:haminic}
\end{align}
where
\begin{equation}
	\mathcal{H}_H = -t\sum_{\ave{\iv,\jv}\!,\s} 
		(c_{\iv\s}^\dagger c_{\jv\s}^{\phantom{\dagger}} + \text{H.c.})
		+U\sum_{\iv} n_{\iv\uparrow} n_{\iv\downarrow} 	
\end{equation}
is the usual Hubbard Hamiltonian describing fermions hopping ($t$ is the hopping integral) between nearest neighbor sites, $\ave{\iv,\jv}$, of a square lattice, $c_{\iv\s}^\dagger$ ($c_{\jv\s}^{\phantom{\dagger}}$) creates (annihilates) a fermion with spin $\s$ at site $\iv$ ($\jv$), $U$ is the strength of the on-site Coulomb repulsion, and  
$n_{\iv\s}\equiv c_{\iv\s}^\dagger c_{\iv\s}^{\phantom{\dagger}}$;
\begin{equation}
	  \mathcal{H}_\text{SO} = V_\text{SO}\sum _{\iv,\s,\s'} \left[i( 
	  	c^{\dagger}_{\iv\s}\s_{\s\!\s'}^x c _{\iv+\widehat{\mathbf{y}}\s'}^{\phantom{\dagger}} 
		- c^{\dagger}_{\iv\s}\s^y_{\s\!\s'}
		c _{\iv+\widehat{\mathbf{x}}\s'}^{\phantom{\dagger}}) + \text{H.c.} 
		\right]
\end{equation}	  
describes the Rashba-type spin-orbit coupling, where $V_\text{SO}$ is the strength of the Rashba SOC, and $\sigma^\tau_{\s\!\s'}$ is the $\s,\s'(=\uparrow,\downarrow)$ element of the Pauli matrices $\widehat{\sigma}^\tau$, $\tau = x,y$; H.c.\ stands for `hermitian conjugate of the previous expression'.
We note that $\mathcal{H}_\text{SO}$ is written in the form of a kinetic term, which means that the \emph{local} Rashba contribution is neglected. 
The reason for this lies in the fact that here we are primarily concerned in highlighting the effects arising from the breakdown of spatial inversion symmetry, which is preserved by the local Rashba SOC~\cite{Mii14}.

We probe the magnetic properties of the model within a Hartree-Fock approximation \cite{Dzierzawa92}, which allows us to decouple the quartic terms of the Hamiltonian, leading to a quadratic form, with the cost of adding effective Weiss fields with respect to which minimization will be sought; see below.
In order to investigate the possibility of stabilizing spiral magnetic phases, we let the magnetization vector be defined as \cite{Dzierzawa92,Costa17a,Costa18a,SOusaJr20}
\begin{equation}
	\mv_\iv = \ave{\Sv_\iv} = m[\cos(\Qv\cdot \Rv),\sin(\Qv\cdot \Rv),0],
\end{equation}
with
\begin{equation}
	\mathbf{S}_{\iv} = \frac{1}{2}\sum_{\s,\s'=\pm} c^{\dagger}_{\iv\s}
		\pmb{\s}_{\s \s'}c_{\iv\s'}^{\phantom{\dagger}}, 
\end{equation}
where $\pmb{\s}$ is the vector operator whose components are the Pauli matrices, and the magnetic wave vector, $\Qv\equiv~(q_x,q_y)$, characterizes the spiral phases. 
Within our scheme, both wave vector and magnetization amplitude, $m$, are determined self-consistently. 

%%%%%%%%%%  Fig. 1  %%%%%%%%%%%
\begin{figure}[t]
\centering
\includegraphics[scale=0.325]{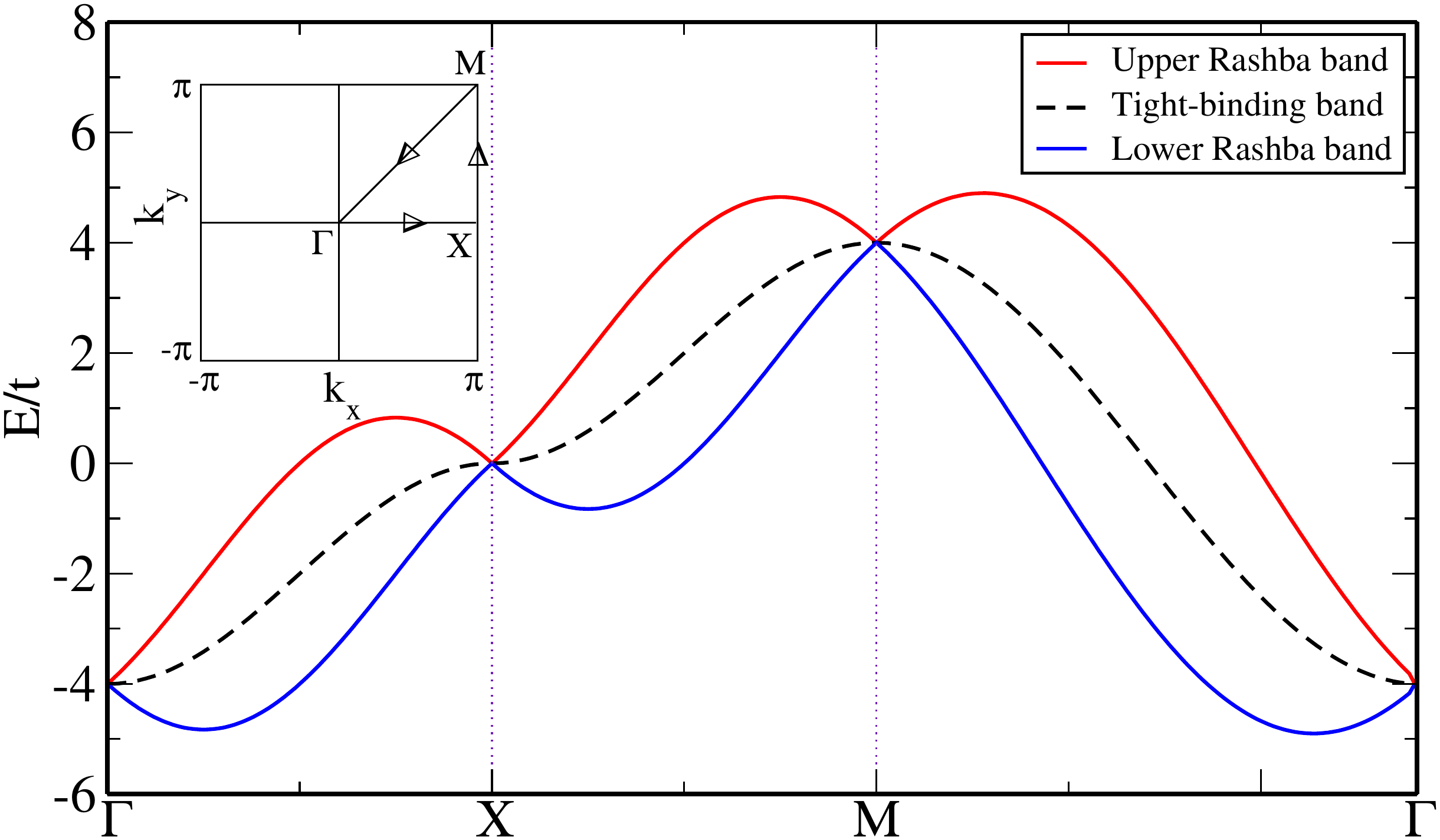} 
\caption{Band structure along a path in $\kv$-space across the high-symmetry points of the square lattice (see inset).
The black dashed curve is $\varepsilon _\kv$, the tight-binding band in the absence of the Rashba SOC, while the upper (red) and lower (blue) solid lines are the Rashba bands for $V_\text{SO}/t=1$, and are labelled by $\alpha=\pm$, respectively. 
}
\label{fig:bandstruc}
\end{figure}
%%%%%%%%%%%%%%%%%%%%%%%%%%%%%%%

The Hartree-Fock Hamiltonian then becomes
\begin{align}
	\label{eq:HFham}
\mathcal{H}_\text{HF} &= -t\sum_{\ave{\iv,\jv},\s} (c_{\iv\s}^\dagger c_{\jv\s}^{\phantom{\dagger}} 
+ c_{\jv\s}^\dagger c_{\iv\s}^{\phantom{\dagger}}) \nonumber\\  
& + V_\text{SO}\sum_{\iv,\s,\s'} \left[i(c^\dagger_{\iv\s}\s^x_{\s\!\s'}
c_{\iv+\widehat{\mathbf{y}}\s'} - c^\dagger_{\iv\s}\s^y_{\s\!\s'}c _{\iv+\widehat{\textbf{x}}\s'}^{\phantom{\dagger}}) + \text{H.c.}\right]\nonumber\\
& + U\sum_{\textbf{i}} \left[ \frac{n}{2}n_\textbf{i} - 2\textbf{m}_{\textbf{i}}\cdot \textbf{S}_{\textbf{i}} - \frac{n^2}{4}  + m^2\right],
\end{align}
where $n$ is the the average electronic density. 
The diagonalization of $\mathcal{H}_\text{HF}$ is more readily worked out in reciprocal space.
Upon Fourier transforming Eq.\,\eqref{eq:HFham}, we obtain,
\begin{align}
	\mathcal{H}_\text{HF} &= \sum_\kv
	(\varepsilon_\kv^\phdag c_{\kv\uparrow}^\dagger c_{\kv\uparrow}^\phdag
	+\varepsilon_{\kv+\qv}^\phdag c_{\kv+\qv\uparrow}^\dagger c_{\kv+\qv\uparrow}^\phdag
	+\frac{1}{2}n Un_\kv)\nonumber\\
		& - Um \sum _\kv (c_{\kv\uparrow}^\dagger c_{\kv+\qv\downarrow}^\phdag
			+c_{\kv+\qv\downarrow}^\dagger c_{\kv\uparrow}^\phdag)\nonumber\\			 
			&+ \sum _{\kv,\s,\s'} (\widehat{V}_\kv)_{\s\s'}^\phdag c_{\kv\s}^\dagger c_{\kv\s'}^\phdag, 
\label{eq:hamkspace}
\end{align}
where 
\begin{equation}
	\widehat{V}_\kv \equiv 2V_\text{SO}(\s^x\sin{k_y} - \s^y\sin{k_x}),
\label{eq:Vhat}	
\end{equation}
and the dispersion relation is $\varepsilon _\kv = -2t(\cos{k_x}+\cos{k_y})$.
In what follows, we perform our analyses on lattices of $200\times200$ sites, which are large enough to disregard finite-size effects.

\subsection{The non-interacting limit}
\label{ssec:Nonintsys}

If we set $U=0$, the original Hamiltonian can be diagonalized by Fourier transforming into $\kv$-space, leading to the bands
\begin{align}
	E_\kv^\pm = \varepsilon _\kv^{\phantom{\pm}} \pm 2V_\text{SO}\sqrt{\sin^2 k_x+\sin^2 k_y},
\label{eq:bands}
\end{align}
and its correspondents eigenvectors,
\begin{align}
	| \pm,\kv \rangle = \dfrac{1}{\sqrt{2}}\left(
	\begin{array}{c}
	1\\
	\pm e^{-i\phi}
	\end{array}
	\right)
\label{eq:eignvctors}
\end{align}
in the spinor basis $\Psi_\kv = ( c_{\kv\uparrow}^\dagger, c_{\kv\downarrow}^\dagger )$, with $\tan\phi = \sin{ky}/\sin{kx}$.
Figure \ref{fig:bandstruc} shows the splitting of the tight-binding band, $ \varepsilon _\kv$, caused by the Rashba SOC:  the lack of spatial inversion symmetry breaks the spin degeneracy in the conventional dispersion relation, thus giving rise to two bands. 
Accordingly, these single-particle bands, each of which labelled by a `chirality', $\al=\pm$, describe a `Rashba Metal' (RM); that is, the spin texture acquires a momentum dependence.   

Examples of the non-interacting density of states with Rashba SOC may be found in Refs.\,\cite{Li11,Ptok18,Hutchinson18}, from which we see that the van Hove singularity (vHS) present at the Fermi energy, $\varepsilon_\text{F}$, at half filling is split in two, symmetrically distributed around $\varepsilon_\text{F}$; also, the nesting of the Fermi surface at half filling is suppressed.

\subsection{The interacting case}
\label{ssec:intsys}

Introducing Nambu spinors, 
\begin{equation}
	\Phi _\kv = (c_{\kv\uparrow}^\dagger, c_{\kv\downarrow}^\dagger,  c_{\kv+\qv\uparrow}^\dagger,
	c_{\kv+\qv\downarrow}^\dagger), 
\end{equation}
the Hartree-Fock Hamiltonian, Eq.\,\eqref{eq:hamkspace}, becomes 
\begin{align}
\label{eq:spinbasis}
\mathcal{H}_\text{HF}  = \frac{1}{2}\sum _\kv \Phi _\kv \widehat{H}_\kv \Phi _\kv^{\dagger}   
+ \sum _\kv\left(\mu n + Um^2 - \frac{1}{4}U n ^2 \right),
\end{align}
where
\begin{eqnarray}
\label{eq:matrix}
	 \widehat{H}_\kv = \left(
		\begin{array}{cccc}
\overline{E}_\kv  & (\widehat{V}_\kv)_{\uparrow \downarrow } & 0 & -Um  \\
(\widehat{V}_\kv)_{ \downarrow \uparrow } & \overline{E}_\kv  & -Um & 0 \\
0 & -Um & \overline{E}_{\kv+\qv}  & (\widehat{V}_{\kv+\qv})_{\uparrow \downarrow} \\
-Um & 0 & (\widehat{V}_{\kv+\qv})_{\downarrow \uparrow } & \overline{E}_{\kv+\qv}, \\
\end{array}
\right)
\end{eqnarray}
with 
\begin{equation}
	\overline{E}_\kv = \varepsilon _\kv -\mu + \frac{1}{2}Un,
\end{equation}
and the $(\widehat{V}_\kv)_{\s \s'}$ are the spin-orbit matrix elements [see Eq.\,\eqref{eq:Vhat}]. 
With the HF Hamiltonian, Eq.\,\eqref{eq:spinbasis}, one solves a Schr\"odinger equation for a given $\kv$, namely  
\begin{equation}
    \widehat{H}_\kv |\psi_\kv^\nu \rangle = \lambda _\kv^{(\nu)}|\psi_\kv^\nu \rangle, 
    \label{eq:Hkpsik}
\end{equation}
from which we extract the eigenstates $|\psi_\kv^\nu \rangle $ and their corresponding eigenvalues $\lambda _\kv^{(\nu)}$, with $\nu$ being an integer labelling the quasiparticle bands. Therefore, the Helmholtz free energy may therefore be written as
\begin{align}
\label{eq:helm}
F =-\frac{1}{\beta} \sum _{\kv,\nu} \ln[1 + e^{-\beta\lambda_\kv^{(\nu)}}]+\mu n 
- U\left(\frac{n^2}{4} - m^2\right),
\end{align}
where $\beta = 1/k_\text{B}T$.

The fields $m$, $\Qv=(q_x,q_y)$, and $\mu$ are determined self-consistently through the minimization of the Helmholtz free energy, 
\begin{align}
\label{eq:minhelm}
	\left \langle \frac{\partial F}{\partial \mu} \right \rangle = \left \langle \frac{\partial F}{\partial m} \right \rangle = \left \langle \frac{\partial F}{\partial  q_x} \right \rangle = \left \langle \frac{\partial F}{\partial  q_y} \right \rangle = 0,
\end{align}
where one should not discard the possibility of having $q_x\neq q_y$.
At this point, it is worth making a technical remark. 
Even though we are ultimately interested in the ground state properties of the model, it turned out that the gain in convergence steps to find the minima is significant if we perform the minimization process at very low temperatures (hence through the free energy) instead of the total (internal) energy. 
The errors involved by working at very low, but finite temperatures are indeed small; for instance, at a temperature $T\sim 1 \times 10^{-4} $ (in units of $t/k_\text{B}$), the difference between the free energy and the ground state energy is smaller than $10^{-6}$ (in units of $t$).

\begin{figure}[t]
    \centering
    \includegraphics[scale=0.35]{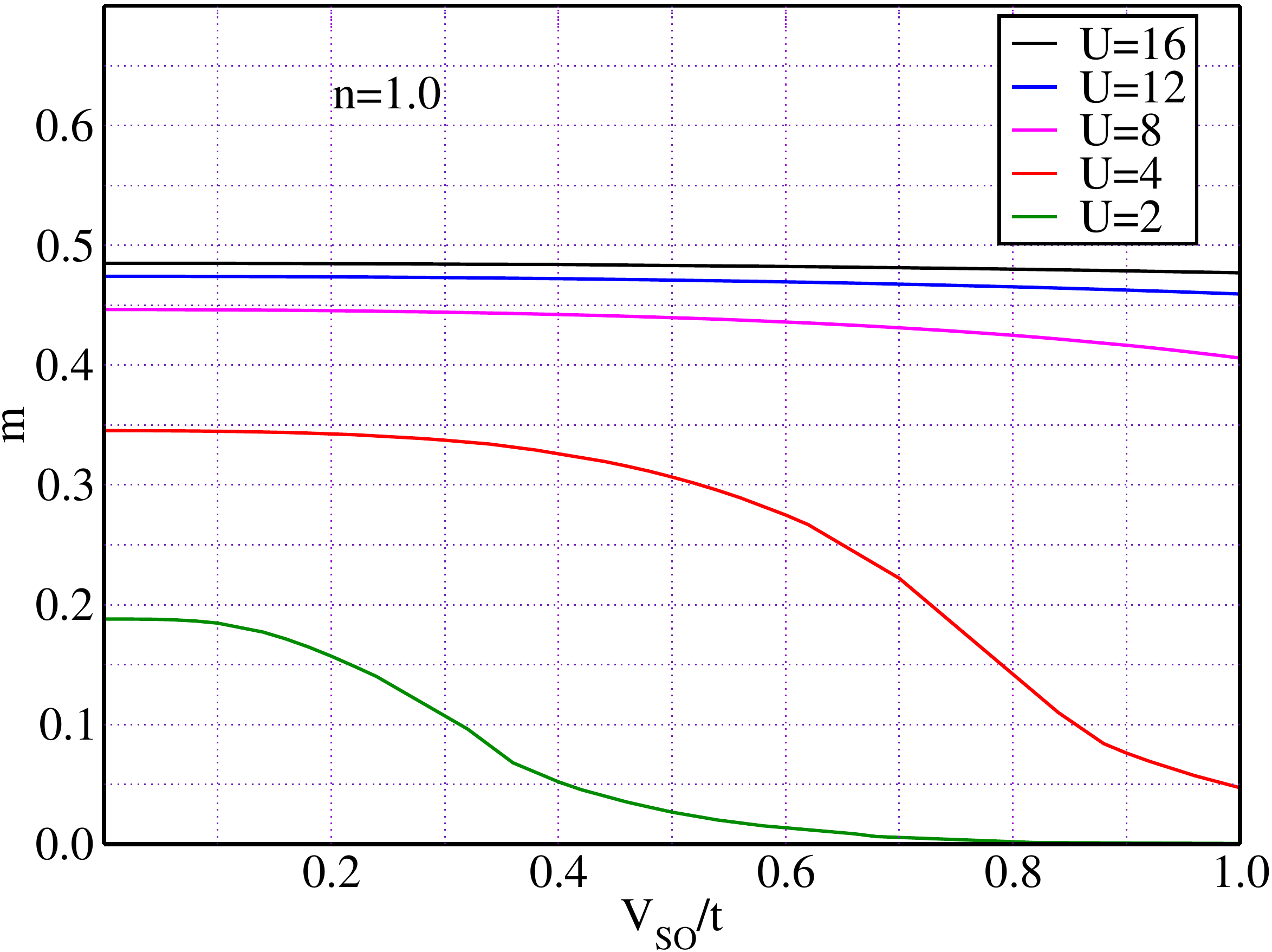}
    \caption{Magnetization amplitude as a function of the spin-orbit coupling strength $V_{SO}$, at half filling, and for different values of $U/t$, which decrease from the top curve to the bottom one.}
    \label{fig:mxvso}
\end{figure}   %Fig 2

Having in mind that several magnetic arrangements may occur, here we adopt the following strategy to determine the ground state. 
For fixed values of $U/t$, $V_\text{SO}$, and band filling, $n$, we calculate the free energy assuming different magnetic wave vectors, $\Qv=(q_x,q_y)$.
For instance, the antiferromagnetic state corresponds to $m\neq 0$ and $q_x=q_y=\pi$, the ferromagnetic state to $m\neq 0$ and $q_x=q_y=0$, and the paramagnetic state to $m=0$. 
We also allow for other phases with $m\neq 0$, such as striped phases with $\Qv=(q, 0)$ [and its symmetric,  $\Qv=(0, q)$], or $\Qv=(q, \pi)$ [$\Qv=(\pi, q)$], as well as general spiral phases, $\Qv=(q, q)$.  
With all the possible lowest free energies at hand, the ground state for the chosen values of $U$, $V_\text{SO}$, and $n$ corresponds to the minimum free energy; the values of $q$ are also varied, and checked whether changes increase or decrease the free energy.
We then repeat for several other values of the control variables to set up the phase diagrams.  
In the next section, we present and discuss the results of this minimization for different values of the electronic density and interaction strengths; we have found instructive to separate the discussion into undoped and doped cases.

\begin{figure}[t]
    \centering
\includegraphics[scale=0.5]{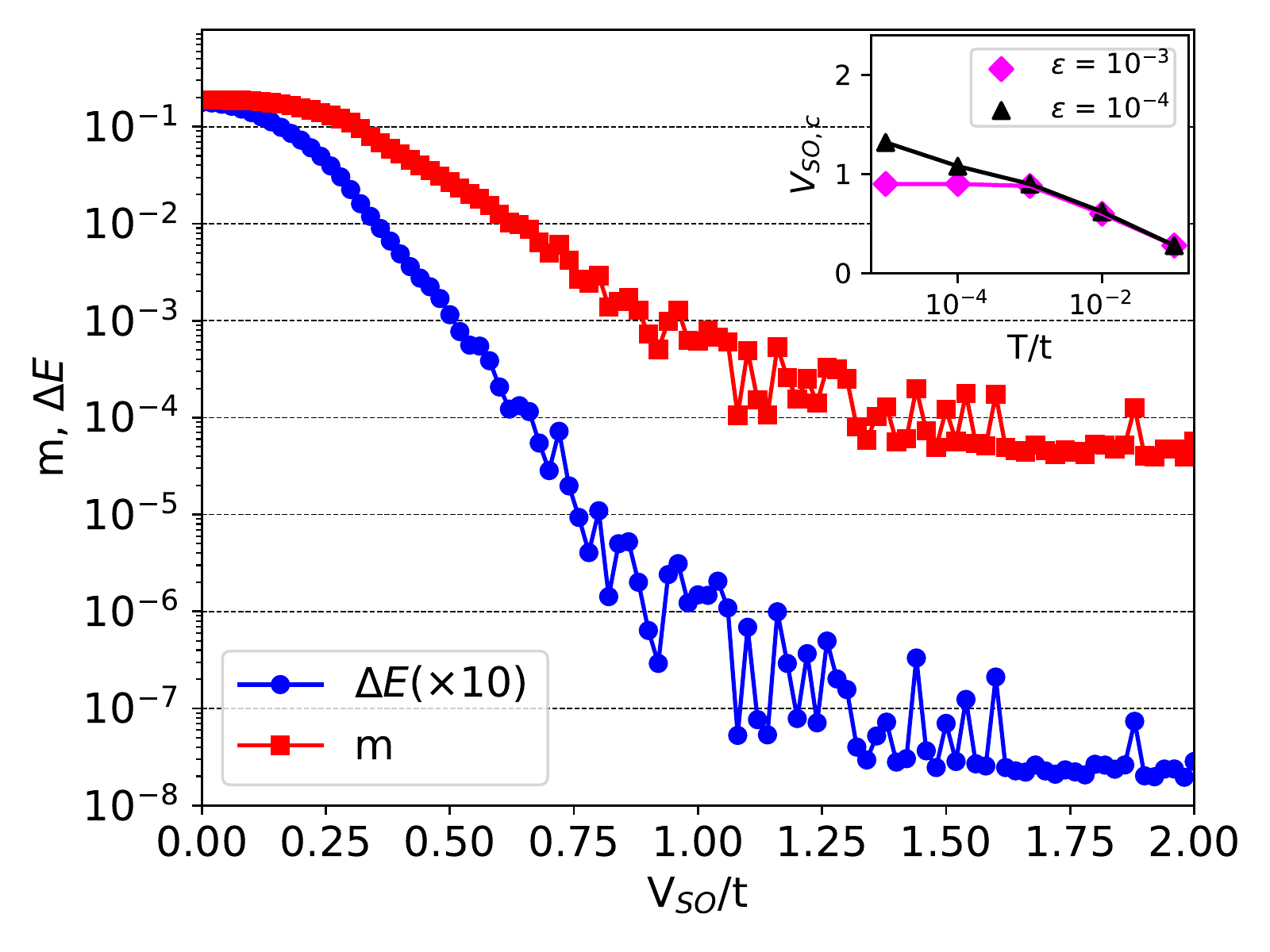}
    \caption{Log-linear plots for the magnetization amplitude, $m$ (red squares), and the energy gap between antiferromagnetic and paramagnetic states, $\Delta E$ (blue circles), as functions of $V_\text{SO}$ for $U/t=2.0$ and $n=1.0$. 
    The inset shows the critical values of the Rashba SOC, $V_{\text{SO},c}$, for two fixed values of the `tolerance as zero', $\varepsilon$, as functions of the temperature used in the minimization process [see Eq.\,\eqref{eq:minhelm}].
    }
   \label{fig:deltaE}
\end{figure}  %Fig 3

\section{Results and discussions}
\label{sec:results}

\subsection{Half-filling}
\label{ssec:HFD}

In the absence of the Rashba SOC, the ground state of the Hubbard model at half filling is antiferromagnetic for any $U>0$, as predicted  both by the present HF approximation \cite{Dzierzawa92,Igoshev15}, and by determinant Quantum Monte Carlo (DQMC) simulations \cite{Hirsch85}.
Further, the magnetization amplitude, $m$, increases with $U$, as a result of the higher degree of fermion localization;
this behavior is indeed verified within our approach when we follow the values of $m$ with $U$ for $V_\text{SO}=0$ in Fig.\,\ref{fig:mxvso}.
Actually, as a combined result of the van Hove singularity and nesting at half filling, $m~\sim~(t/U)~\exp  [-2\pi\sqrt{t/U}]$ within a HF approximation~\cite{Hirsch85}, in the weak coupling regime.

When $V_\text{SO}\neq 0$, spin-flip processes come into play, which tend to disrupt the antiferromagnetically ordered state; this effect should be more effective at smaller on-site couplings.
Indeed, Fig.\,\ref{fig:mxvso} shows that the magnetization amplitude, $m$, decreases faster with $V_\text{SO}$ for the smaller strengths of the Coulomb repulsion, and that for  $U/t \gtrsim 8$ the SOC hardly affects the antiferromagnetic ordering within the physically appealing range of values $V_\text{SO}/t\lesssim 1$.

\begin{figure}[t]
    \centering
    \includegraphics[scale=0.5]{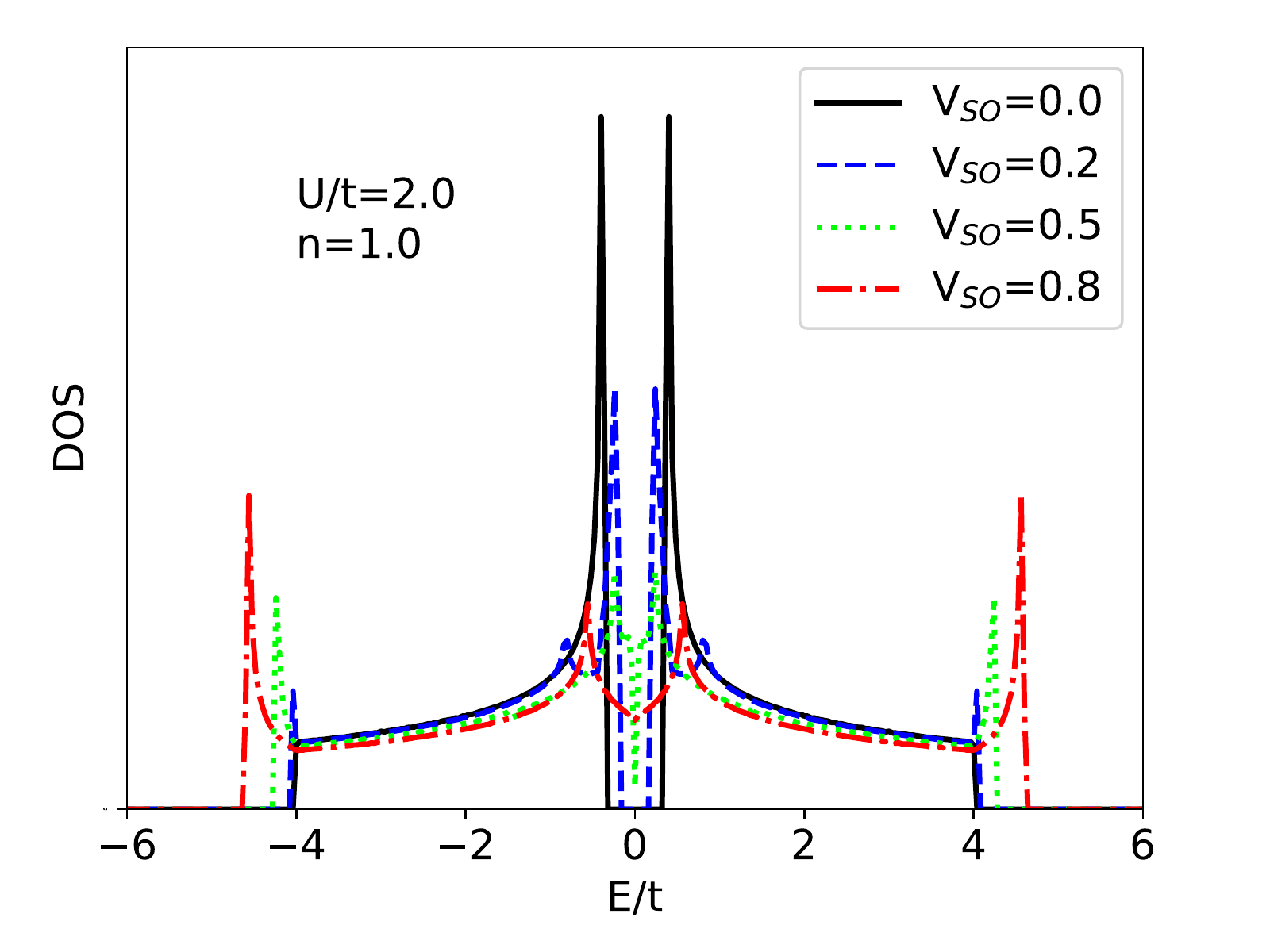}
    \caption{Density of states at half filling with $U/t=2$, and different values of $V_{SO}/t$.
    }
    \label{fig:DOS-MIT}. % Fig 4
\end{figure}

Figure \ref{fig:mxvso} also shows that the magnetization amplitude, $m$, approaches zero very slowly with $V_\text{SO}$.
Let us then take a closer look at the results for  $U/t=2$, for which the behavior of $m$ with $V_\text{SO}/t$ is shown again in Fig.\,\ref{fig:deltaE}, but now as a log-linear plot. 
As mentioned before (in the context of $U=0$), the presence of a Rashba SOC splits the vHS into two peaks; Figure \ref{fig:DOS-MIT} shows that this feature still occurs when $U/t=2$ if $V_\text{SO}\lesssim 0.5$, and that it is located very close to the Fermi energy.  
We may therefore attribute this exponential decay of $m$ to the influence of the nearby vHS-like peak.
As seen for the periodic Anderson model \cite{Hu17}, one expects that the $m\times V_\text{SO}$ curves actually sharpen and acquire the usual order parameter behavior as the lattice size increases, or, equivalently, if the grid used in $\kv$-space gets denser.
Instead of pursuing an investigation along these lines, for our purposes here it suffices to accept as the critical value, $V_{\text{SO},c}$, the value which renders $m/m_0\lesssim\varepsilon$, where $m_0=1/2$ ($\hbar \equiv 1$) is the saturation value for spin-$1/2$, and $\varepsilon$ is a tolerance.
This, in turn, must be balanced with the fact that we are minimizing the free energy at low, but finite temperatures.
Indeed, the inset of Fig.\,\ref{fig:deltaE} shows the estimates of $V_{\text{SO},c}/t$ as a function of temperature for both $\varepsilon=10^{-3}$ and $\varepsilon=10^{-4}$: we see that the choice $\varepsilon=10^{-3}$ leads to a stable value of $V_{\text{SO},c}/t\approx 0.9$ as the temperature is lowered, so we take this as our working tolerance.

\begin{figure}[t]
    \centering
    \includegraphics[scale=0.5]{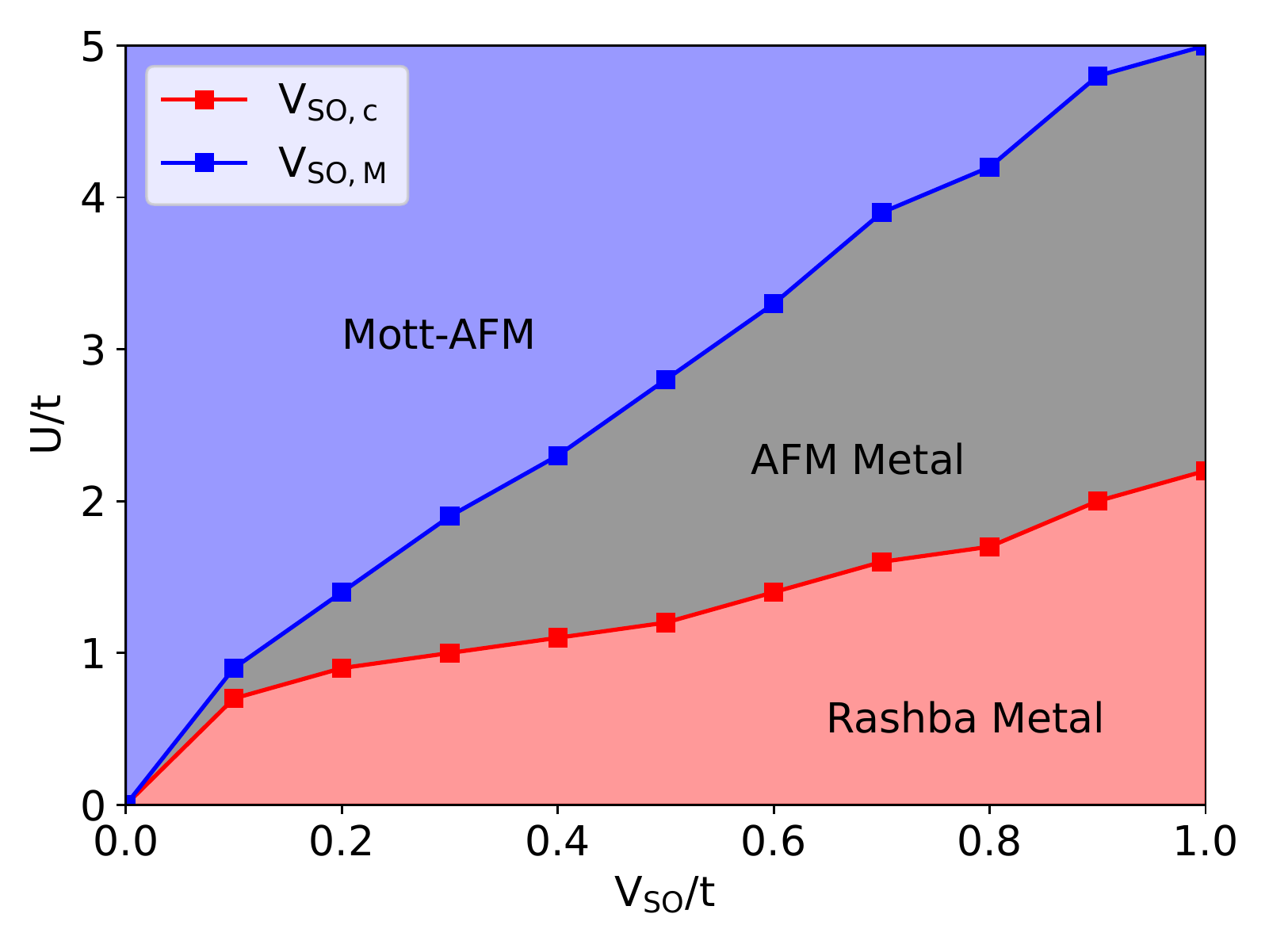}
    \caption{Hartree-Fock phase diagram at half filling. 
    $V_{\text{SO},c}$ denotes the critical curve for the transition between Rashba and antiferromagnetic metals, while $V_{\text{SO},M}$ denotes the critical curve for the metal-insulator transition between  antiferromagnetic states. 
    }
    \label{fig:diaghf}
\end{figure}

A question which immediately arises is whether the suppression of the antiferromagnetic phase is accompanied by a transition to a metallic state.
In order to investigate this issue, we examine the spectral properties. 
On the one hand, we calculate the total (ground state) energy assuming an AFM state as well as a RM state, and observe the gap, $\Delta E$, between these energies as a function of $V_\text{SO}/t$, for fixed $U/t$; the outcome for $U/t=2$ is shown in Fig.\,\ref{fig:deltaE}. 
One finds a decay towards zero faster than the one for $m$: 
more specifically, $\Delta E$ reaches a tolerance of $\tilde{\varepsilon}/t\sim 10^{-3}$ at $V_{\text{SO},M}/t\approx 0.5$, which is almost half of $V_{\text{SO},c}/t$, as determined from the behavior of $m$.   
We have also calculated the density of states (DOS) on both sides of the transition; data for $U/t=2$ are shown in Fig.\,\ref{fig:DOS-MIT}. 
For small values of $V_\text{SO}$ the behavior is reminiscent of what one would expect for a Mott insulator, namely the van Hove singularity is split into two peaks separated by a gap around the Fermi energy. 
As $V_\text{SO}$ increases, the gap narrows, forming a pseudo-gap near $V_{\text{SO},M}/t\approx 0.5$; upon further increase of $V_\text{SO}$ the gap closes leading to full metallic behavior.
As a final test of our findings, we examined the possibility of the difference between the two critical points being caused by very shallow minima of the free energy, but found that the minima are safely separated in $V_\text{SO}$.
We therefore conclude that within our approach we have found a sequence of two transitions as $V_\text{SO}$ increases, namely first a Mott-AFM to an AFM Metal, and then to a Rashba Metal.

By employing the same procedure for different values of $V_\text{SO}$, we obtain the phase diagram shown in Fig.\,\ref{fig:diaghf}.
We see that increasing $V_\text{SO}$ at fixed $U$ first drives the system from a Mott Antiferromagnetic (AFM) phase to a metallic AFM phase; then, as $U$ is further increased, the AFM gives way to a Rashba metallic phase.  
The fact that a Rashba SOC favors the appearance of a metallic phase is in line with the findings of Ref.\,\onlinecite{Brosco20}, in which the strong $V_\text{SO}$  regime was considered.

%!!!!!!!!!!!!!!!!!!!!!!!!!!!!!!!!!!!!!!!!!!!!!!!!!!!!!!!!!!!!

\begin{figure}[t]
    \centering
    \includegraphics[scale=0.3]{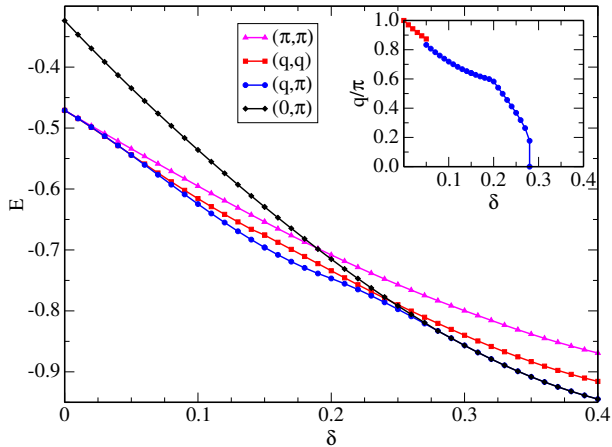}
    \caption{Energy for different phases as a function of doping, $\delta$, for fixed $V_\text{SO}=0.5$ and $U/t = 9.0$. 
    The inset shows the value of $q/\pi$ as a function of doping for the striped phase, $(q,\pi)$.}
    \label{fig:energy_delta} %Fig 5
\end{figure}

\subsection{Doped regime}
\label{ssec:ADR}

An early ground state HF phase diagram, $U$ vs.\ doping, $\delta\equiv1-n$, for the square lattice Hubbard model only involved the possibility of  FM, AFM (N\'eel) and paramagnetic phases \cite{Hirsch85}; as generally expected for mean-field--like approaches, the outcome was qualitatively similar to the phase diagram obtained for the cubic lattice~\cite{Penn66}.  
Later on, a HF approach allowing for spiral phases revealed that the N\'eel phase is indeed unstable against finite doping, giving way to incommensurate magnetic arrangements, as well as to striped phases  \cite{Dzierzawa92,Igoshev15}.
The suppression of antiferromagnetism for any doping is in agreement with predictions from DQMC simulations~\cite{Hirsch85}, thus adding extra reliability to this approach. 
Indeed, this approach has revealed myriads of magnetic phases both in the Kondo-lattice model \cite{Costa17a}, and in a model \cite{Bertussi09} for coexistence of magnetism and superconductivity \cite{Costa18a}; the latter work describes some features of the magnetic arrangements tuned by doping, as experimentally observed in the borocarbide family of materials \cite{ElMassalami12,ElMassalami13,ElMassalami14}.

\begin{figure}[t]
    \centering
    \includegraphics[scale=0.4]{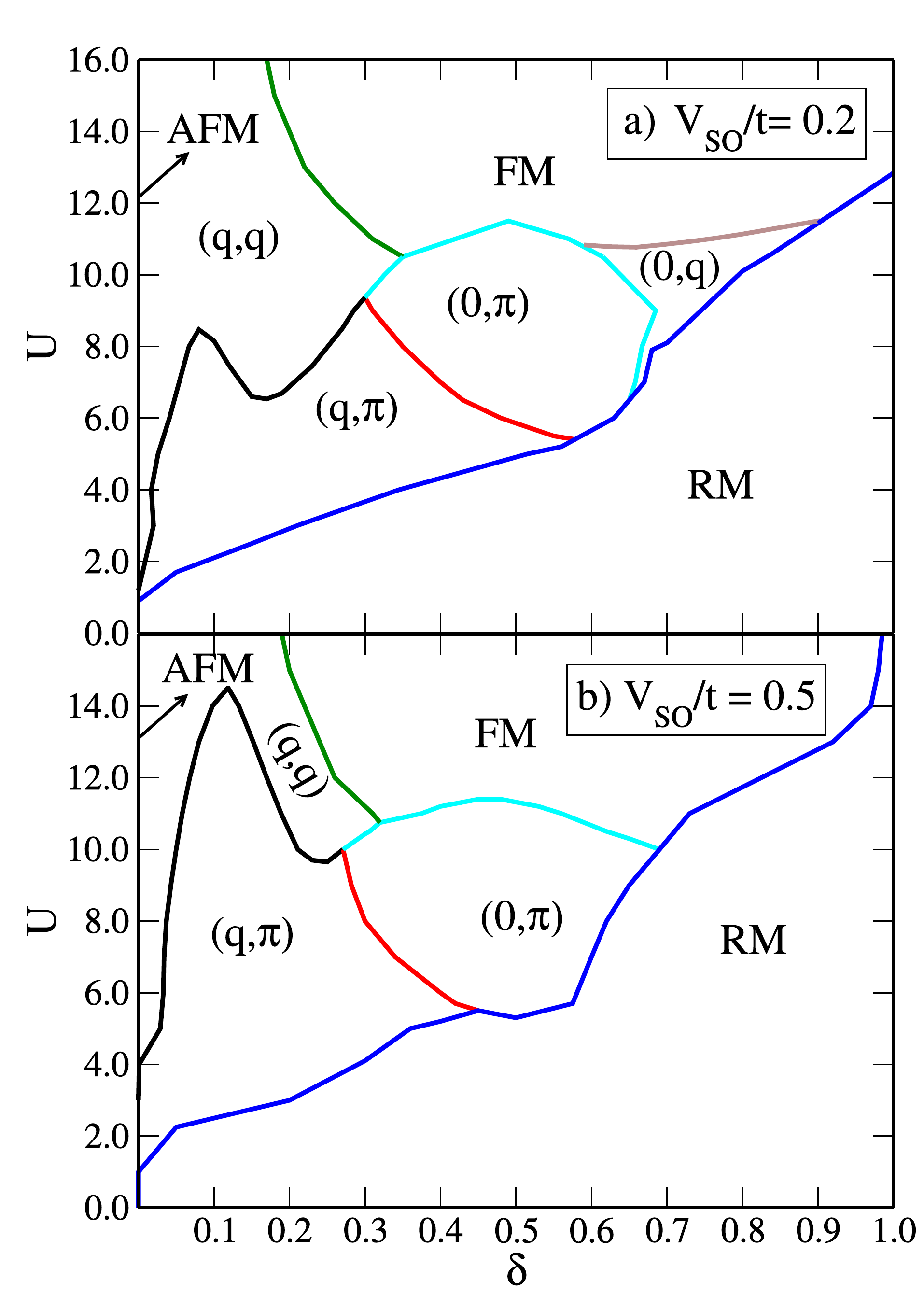}
	\caption{Hartree-Fock phase diagrams $U$ vs.\ doping, $\delta$, for (a) $V_\text{SO}/t=0.2$, and (b)  $V_\text{SO}/t=0.5$. AFM stands for antiferromagnetic, FM for ferromagnetic, and RM for Rashba metal, while the spiral phases are labelled by $\Qv\equiv (q_x,q_y)$; see text.
	The black dashed lines separate the insulating phase, predominantly in the low-doping regime from the metallic phase in the high-doping regime.}	 
	\label{fig:Udelta}
\end{figure}

Away from half filling, it is illustrative to compare the lowest energies, assuming different magnetic states, as functions of doping, as shown in Fig.\,\ref{fig:energy_delta} for $V_\text{SO}/t=0.5$ and $U/t=9$.
At very low doping, $\delta \lesssim 0.05$, the lowest energy state corresponds to $\Qv=(q,q)$, with $q$ decreasing from $\pi$ roughly linearly with $\delta$ (red squares in the inset of Fig.\,\ref{fig:energy_delta}).
In the interval $\delta \in [0.05,0.3]$, the ground state corresponds to $\Qv=(q,\pi)$, with $q$ decreasing from 0.84 to 0, as shown in the inset of Fig.\,\ref{fig:energy_delta}.
At this doping the energy for $\Qv=(q,\pi)$ (blue circles) joins smoothly with the one corresponding to $(0,\pi)$ (black diamonds).
By repeating this procedure for other values of $U$ and $V_\text{SO}$, we generate the phase diagrams displayed in Fig.\,\ref{fig:Udelta}.

\begin{figure*}[t]
    \centering
    \includegraphics[scale=0.8]{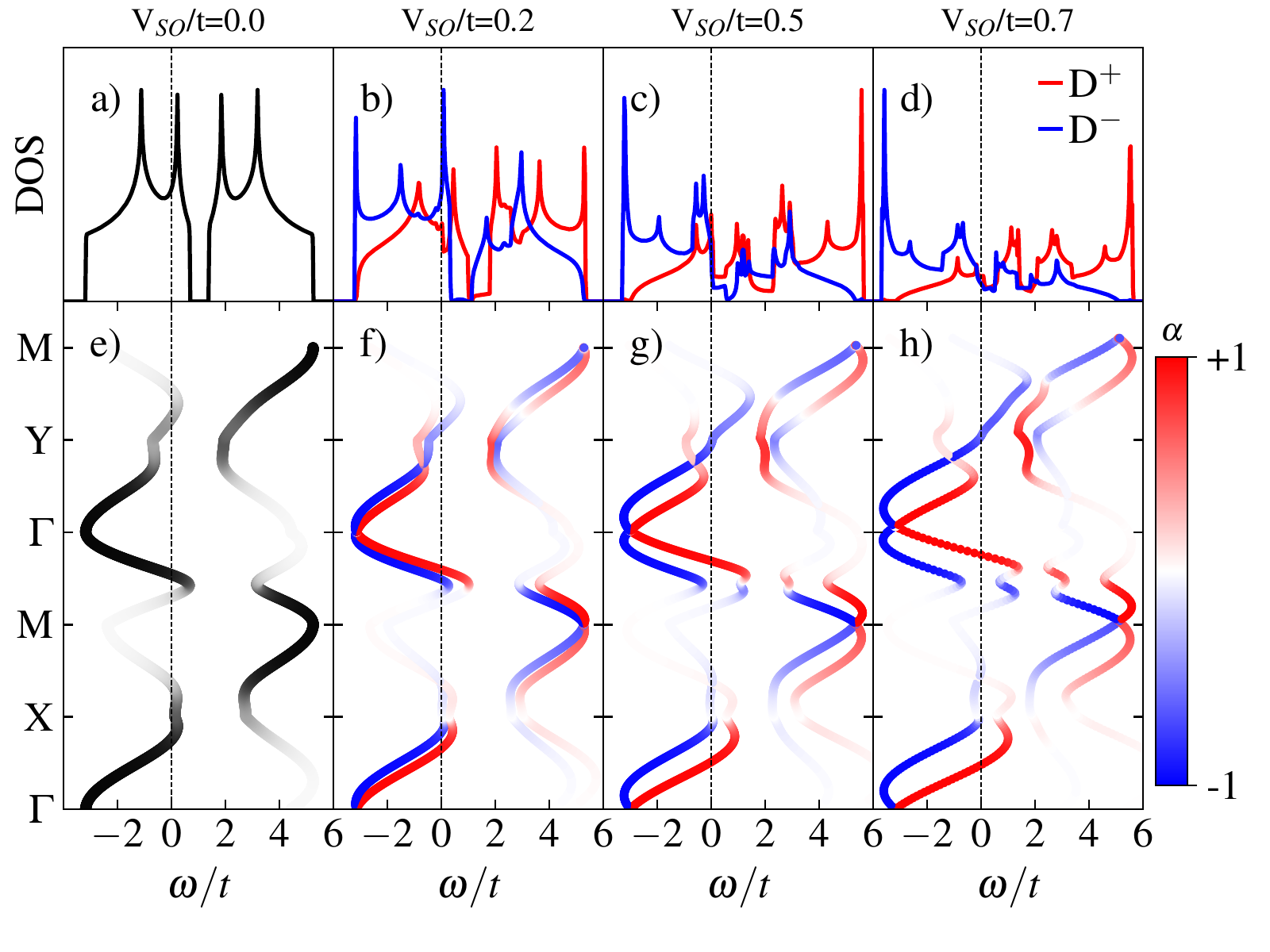}
    \caption{Upper panel: Density of states for $n=0.85$, $U/t=5$ and $V_{SO}=$ 0.0 a), 0.2 b), 0.5 c), and 0.7 d). Lower panel: Projected spectral function for $n=0.85$, $U/t=5$ and $V_{SO}=$ 0.0 e), 0.2 f), 0.5 g), and 0.7 h); the chirality $\alpha$ is indicated by the color map. In each case, the Fermi level is indicated by the dashed lines.
    }
    \label{fig:dosvsvso}
\end{figure*}

Several features stand out when Fig.\,\ref{fig:Udelta} is compared with the case $V_\text{SO}=0$ \cite{Dzierzawa92,Liu94,Igoshev15}.
First, for small doping the region of stability for the symmetric arrangement $\Qv=(q,q)$ shrinks as $V_\text{SO}$ increases, in favor of the striped phase $\Qv=(q,\pi)$
\footnote{One should keep in mind that the exchange symmetry $q_x \leftrightarrow q_y$ applies, \textit{i.e.}\ the solutions $\textbf{q}=(q_x,q_y)$ and $\textbf{q}=(q_y,q_x)$ are degenerate.}.
Another noteworthy change is the fact that the phase $\Qv=(0,q)$, located near $\delta=0.7$, disappears with increasing $V_\text{SO}$, so that one has a direct transition between a RM and a ferromagnet at large doping.
By contrast, the size of the striped region, $\Qv=(0,\pi)$, centered around quarter filling and with $U/t$ around 9, is quite insensitive to the magnitude of the SOC.

From the above findings, we may conclude that, at least for dopings up to quarter filling, SOC tends to favor striped magnetic arrangements.
Indeed, within a semiclassical picture the Rashba SOC favors the spins to point along directions perpendicular to the hopping direction; thus, a delicate balance with the Pauli principle is achieved by fermions flowing along the the $x$ and $y$ lattice directions.

In order to understand the evolution of the magnetic wave vector with $V_\text{SO}$, we first define a spin chirality as~\cite{Park2013},
\begin{equation}
    \alpha (\kv,\nu) =  \langle \psi_\kv^\nu | 2 \textbf{S} \cdot (\hat{\kv} \times \hat{\mathbf{z}} )|\psi_\kv^\nu \rangle,
\end{equation}
where the $|\psi_\kv^\nu \rangle $ are the HF ground states for $U\neq 0$; see Eq.\,\eqref{eq:Hkpsik}. 
When $U\to 0$, $|\psi_\kv^\nu \rangle \to |\pm,\kv\rangle$, as given by Eq.~\eqref{eq:eignvctors}, so that $\alpha (\kv,\nu)\to \pm1$; note, however, that for $U\neq0$, $\alpha (\kv,\nu)$ may also vary continuously between $\pm1$, as we will see below.  
We now define projected spectral functions, as follows:  
\begin{align}
    A^{\pm}(\kv,\omega) = \sum_{\nu} \alpha (\kv,\nu) \delta (\omega - \lambda_\kv^{(\nu)}),
\end{align}
so that the $\delta$-functions are weighted by the spin chirality, whose sign, in turn, labels the superscripts $\pm$ in $A^\pm(\kv,\omega)$. 
With $A^{\pm}(\kv,\omega)$ at hand, we obtain the projected density of states as $D^\pm(\omega) = \sum_\kv A^\pm(\kv,\omega) $.

\begin{figure}[t]
    \centering
    \includegraphics[scale=0.5]{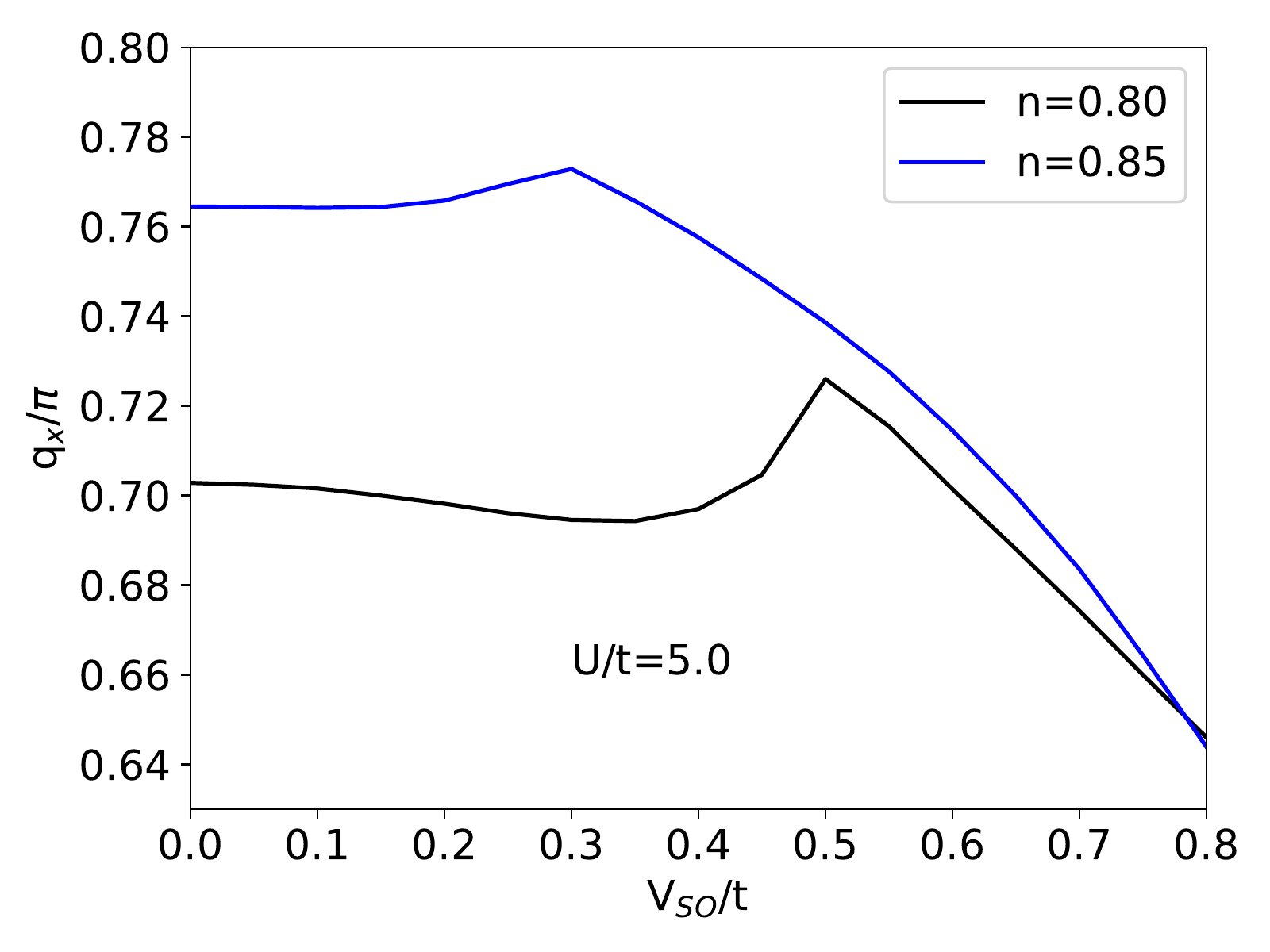}
    \caption{The dependence of the wave vector with $V_\text{SO}$ for $U/t=5.0$.
    }
    \label{fig:qxv} % Fig 8
\end{figure}

The top (bottom) row of Figure \ref{fig:dosvsvso} shows the DOS, $D^\pm$ (spectral function, $A^\pm$), for increasing intensities of the SOC.   
Note that for all values of $V_\text{SO}$, the DOS at the Fermi energy is finite, indicating metallic behavior, as expected.
For $V_\text{SO}/t=0$, the bands are unsplit, as they should, but non-zero values of $V_\text{SO}$ cause a chirality splitting of the bands, with dspersionless features around some points (notably around X) in the Brillouin zone being preserved. 
With increasing $V_\text{SO}$, the system becomes helically polarized, i.e.\ states below the Fermi level acquire a dominant $\al<0$ character.

We are now in a position to examine the evolution of $q$ [in $\Qv=(q,\pi)$] with the SOC intensity. 
Figure~\ref{fig:qxv} shows that for $U/t=5$, and for both $n=0.8$ and $n=0.85$, $q$ slowly changes with $V_\text{SO}$, until it reaches a peak, and then steadily declines. 
Let us focus on Fig.\,\ref{fig:dosvsvso}, corresponding to filling $n=0.85$. 
For $V_\text{SO}/t=0.2$ the Fermi level lies at the vH singularity for $D^-$, so that many states contribute to the averages; by contrast, beyond $V_\text{SO}/t=0.5$, the contribution from the vH singularity is strongly suppressed, and $q$ decreases.  
The inescapable conclusion is that the SOC strongly influences the magnetic ordering through changes in the the spectral weight, especially near the Fermi level.

\section{Conclusions}
\label{sec:concl}

We have considered the Hubbard model in the presence of a Rashba spin-orbit coupling (SOC).
Through a Hartree-Fock (HF) approach which allows for the presence of spiral magnetic arrangements, we have determined ground state phase diagrams in the parameter space of on-site repulsion, $U$, SOC strength, $V_\text{SO}$, and doping, $\delta$.
At half filling, we have established that for fixed $U$ an increasing Rashba SOC drives a sequence of two transitions: from a Mott insulator to a metallic antiferromagnet, and then to a paramagnetic Rashba metal.
In the doped regime, several magnetic phases appear, including ferromagnetic and striped phases. 
Given that one has a very fine control of the repulsive interaction in ultracold atoms, we may envisage the possibility of generating a wide variety of magnetic arrangements simply by varying $U$ and the doping level, say for fixed SOC. 
One may also expect that the HF phase diagrams we have obtained here for a  square lattice, can serve as a guide to three-dimensional systems.

%%%%%%%%%%%%%%%%%%%%%%%%%%%%%%%%%%%%%%%%%%%%%%%%%%%%%%%%%%%%%%%%%%
\section*{ACKNOWLEDGMENTS} 
The authors are grateful to the Brazilian Agencies National Council for Scientific and Technological Development (CNPq), National Council for the Improvement of Higher Education (CAPES), and FAPERJ for funding this project.
N.C.C. acknowledges financial support from the Brazilian Agency CNPq, grant number 313065/2021-7.

%%%%%%%%%%%%%%%%%%%%%%%%%%%%%%%%%%%%%%%%%%%%%%%%%%%%%%%%%%%%%%%%%%

\bibliography{soc_hub}

\begin{thebibliography}{30}%
\makeatletter
\providecommand \@ifxundefined [1]{%
 \@ifx{#1\undefined}
}%
\providecommand \@ifnum [1]{%
 \ifnum #1\expandafter \@firstoftwo
 \else \expandafter \@secondoftwo
 \fi
}%
\providecommand \@ifx [1]{%
 \ifx #1\expandafter \@firstoftwo
 \else \expandafter \@secondoftwo
 \fi
}%
\providecommand \natexlab [1]{#1}%
\providecommand \enquote  [1]{``#1''}%
\providecommand \bibnamefont  [1]{#1}%
\providecommand \bibfnamefont [1]{#1}%
\providecommand \citenamefont [1]{#1}%
\providecommand \href@noop [0]{\@secondoftwo}%
\providecommand \href [0]{\begingroup \@sanitize@url \@href}%
\providecommand \@href[1]{\@@startlink{#1}\@@href}%
\providecommand \@@href[1]{\endgroup#1\@@endlink}%
\providecommand \@sanitize@url [0]{\catcode `\\12\catcode `\$12\catcode
  `\&12\catcode `\#12\catcode `\^12\catcode `\_12\catcode `\%12\relax}%
\providecommand \@@startlink[1]{}%
\providecommand \@@endlink[0]{}%
\providecommand \url  [0]{\begingroup\@sanitize@url \@url }%
\providecommand \@url [1]{\endgroup\@href {#1}{\urlprefix }}%
\providecommand \urlprefix  [0]{URL }%
\providecommand \Eprint [0]{\href }%
\providecommand \doibase [0]{http://dx.doi.org/}%
\providecommand \selectlanguage [0]{\@gobble}%
\providecommand \bibinfo  [0]{\@secondoftwo}%
\providecommand \bibfield  [0]{\@secondoftwo}%
\providecommand \translation [1]{[#1]}%
\providecommand \BibitemOpen [0]{}%
\providecommand \bibitemStop [0]{}%
\providecommand \bibitemNoStop [0]{.\EOS\space}%
\providecommand \EOS [0]{\spacefactor3000\relax}%
\providecommand \BibitemShut  [1]{\csname bibitem#1\endcsname}%
\let\auto@bib@innerbib\@empty
%</preamble>
\bibitem [{\citenamefont {Witczak-Krempa}\ \emph {et~al.}(2014)\citenamefont
  {Witczak-Krempa}, \citenamefont {Chen}, \citenamefont {Kim},\ and\
  \citenamefont {Balents}}]{Witczak14}%
  \BibitemOpen
  \bibfield  {author} {\bibinfo {author} {\bibfnamefont {William}\ \bibnamefont
  {Witczak-Krempa}}, \bibinfo {author} {\bibfnamefont {Gang}\ \bibnamefont
  {Chen}}, \bibinfo {author} {\bibfnamefont {Yong~Baek}\ \bibnamefont {Kim}}, \
  and\ \bibinfo {author} {\bibfnamefont {Leon}\ \bibnamefont {Balents}},\
  }\bibfield  {title} {\enquote {\bibinfo {title} {Correlated quantum phenomena
  in the strong spin-orbit regime},}\ }\href {\doibase
  10.1146/annurev-conmatphys-020911-125138} {\bibfield  {journal} {\bibinfo
  {journal} {Annual Review of Condensed Matter Physics}\ }\textbf {\bibinfo
  {volume} {5}},\ \bibinfo {pages} {57--82} (\bibinfo {year}
  {2014})}\BibitemShut {NoStop}%
\bibitem [{\citenamefont {Bercioux}\ and\ \citenamefont
  {Lucignano}(2015)}]{Bercioux15}%
  \BibitemOpen
  \bibfield  {author} {\bibinfo {author} {\bibfnamefont {Dario}\ \bibnamefont
  {Bercioux}}\ and\ \bibinfo {author} {\bibfnamefont {Procolo}\ \bibnamefont
  {Lucignano}},\ }\bibfield  {title} {\enquote {\bibinfo {title} {Quantum
  transport in {Rashba} spin{\textendash}orbit materials: a review},}\ }\href
  {\doibase 10.1088/0034-4885/78/10/106001} {\bibfield  {journal} {\bibinfo
  {journal} {Reports on Progress in Physics}\ }\textbf {\bibinfo {volume}
  {78}},\ \bibinfo {pages} {106001} (\bibinfo {year} {2015})}\BibitemShut
  {NoStop}%
\bibitem [{\citenamefont {Manchon}\ \emph {et~al.}(2015)\citenamefont
  {Manchon}, \citenamefont {Koo}, \citenamefont {Nitta}, \citenamefont
  {Frolov},\ and\ \citenamefont {Duine}}]{Manchon15}%
  \BibitemOpen
  \bibfield  {author} {\bibinfo {author} {\bibfnamefont {A.}~\bibnamefont
  {Manchon}}, \bibinfo {author} {\bibfnamefont {H.~C.}\ \bibnamefont {Koo}},
  \bibinfo {author} {\bibfnamefont {J.}~\bibnamefont {Nitta}}, \bibinfo
  {author} {\bibfnamefont {S.~M.}\ \bibnamefont {Frolov}}, \ and\ \bibinfo
  {author} {\bibfnamefont {R.~A.}\ \bibnamefont {Duine}},\ }\bibfield  {title}
  {\enquote {\bibinfo {title} {New perspectives for {R}ashba spin-orbit
  coupling},}\ }\href {\doibase 10.1038/nmat4360} {\bibfield  {journal}
  {\bibinfo  {journal} {Nature Materials}\ }\textbf {\bibinfo {volume} {14}},\
  \bibinfo {pages} {871--882} (\bibinfo {year} {2015})}\BibitemShut {NoStop}%
\bibitem [{\citenamefont {Bertinshaw}\ \emph {et~al.}(2019)\citenamefont
  {Bertinshaw}, \citenamefont {Kim}, \citenamefont {Khaliullin},\ and\
  \citenamefont {Kim}}]{Bertinshaw19}%
  \BibitemOpen
  \bibfield  {author} {\bibinfo {author} {\bibfnamefont {Joel}\ \bibnamefont
  {Bertinshaw}}, \bibinfo {author} {\bibfnamefont {Y.K.}\ \bibnamefont {Kim}},
  \bibinfo {author} {\bibfnamefont {Giniyat}\ \bibnamefont {Khaliullin}}, \
  and\ \bibinfo {author} {\bibfnamefont {B.J.}\ \bibnamefont {Kim}},\
  }\bibfield  {title} {\enquote {\bibinfo {title} {Square lattice iridates},}\
  }\href {\doibase 10.1146/annurev-conmatphys-031218-013113} {\bibfield
  {journal} {\bibinfo  {journal} {Annual Review of Condensed Matter Physics}\
  }\textbf {\bibinfo {volume} {10}},\ \bibinfo {pages} {315--336} (\bibinfo
  {year} {2019})}\BibitemShut {NoStop}%
\bibitem [{\citenamefont {Gotlieb}\ \emph {et~al.}(2018)\citenamefont
  {Gotlieb}, \citenamefont {Lin}, \citenamefont {Serbyn}, \citenamefont
  {Zhang}, \citenamefont {Smallwood}, \citenamefont {Jozwiak}, \citenamefont
  {Eisaki}, \citenamefont {Hussain}, \citenamefont {Vishwanath},\ and\
  \citenamefont {Lanzara}}]{Gotlieb18}%
  \BibitemOpen
  \bibfield  {author} {\bibinfo {author} {\bibfnamefont {Kenneth}\ \bibnamefont
  {Gotlieb}}, \bibinfo {author} {\bibfnamefont {Chiu-Yun}\ \bibnamefont {Lin}},
  \bibinfo {author} {\bibfnamefont {Maksym}\ \bibnamefont {Serbyn}}, \bibinfo
  {author} {\bibfnamefont {Wentao}\ \bibnamefont {Zhang}}, \bibinfo {author}
  {\bibfnamefont {Christopher~L.}\ \bibnamefont {Smallwood}}, \bibinfo {author}
  {\bibfnamefont {Christopher}\ \bibnamefont {Jozwiak}}, \bibinfo {author}
  {\bibfnamefont {Hiroshi}\ \bibnamefont {Eisaki}}, \bibinfo {author}
  {\bibfnamefont {Zahid}\ \bibnamefont {Hussain}}, \bibinfo {author}
  {\bibfnamefont {Ashvin}\ \bibnamefont {Vishwanath}}, \ and\ \bibinfo {author}
  {\bibfnamefont {Alessandra}\ \bibnamefont {Lanzara}},\ }\bibfield  {title}
  {\enquote {\bibinfo {title} {Revealing hidden spin-momentum locking in a
  high-temperature cuprate superconductor},}\ }\href {\doibase
  10.1126/science.aao0980} {\bibfield  {journal} {\bibinfo  {journal}
  {Science}\ }\textbf {\bibinfo {volume} {362}},\ \bibinfo {pages} {1271--1275}
  (\bibinfo {year} {2018})}\BibitemShut {NoStop}%
\bibitem [{\citenamefont {Kolkowitz}\ \emph {et~al.}(2017)\citenamefont
  {Kolkowitz}, \citenamefont {Bromley}, \citenamefont {Bothwell}, \citenamefont
  {Wall}, \citenamefont {Marti}, \citenamefont {Koller}, \citenamefont {Zhang},
  \citenamefont {Rey},\ and\ \citenamefont {Ye}}]{Kolkowitz17}%
  \BibitemOpen
  \bibfield  {author} {\bibinfo {author} {\bibfnamefont {S.}~\bibnamefont
  {Kolkowitz}}, \bibinfo {author} {\bibfnamefont {S.~L.}\ \bibnamefont
  {Bromley}}, \bibinfo {author} {\bibfnamefont {T.}~\bibnamefont {Bothwell}},
  \bibinfo {author} {\bibfnamefont {M.~L.}\ \bibnamefont {Wall}}, \bibinfo
  {author} {\bibfnamefont {G.~E.}\ \bibnamefont {Marti}}, \bibinfo {author}
  {\bibfnamefont {A.~P.}\ \bibnamefont {Koller}}, \bibinfo {author}
  {\bibfnamefont {X.}~\bibnamefont {Zhang}}, \bibinfo {author} {\bibfnamefont
  {A.~M.}\ \bibnamefont {Rey}}, \ and\ \bibinfo {author} {\bibfnamefont
  {J.}~\bibnamefont {Ye}},\ }\bibfield  {title} {\enquote {\bibinfo {title}
  {Spin--orbit-coupled fermions in an optical lattice clock},}\ }\href
  {\doibase 10.1038/nature20811} {\bibfield  {journal} {\bibinfo  {journal}
  {Nature}\ }\textbf {\bibinfo {volume} {542}},\ \bibinfo {pages} {66--70}
  (\bibinfo {year} {2017})}\BibitemShut {NoStop}%
\bibitem [{\citenamefont {Song}\ \emph {et~al.}(2019)\citenamefont {Song},
  \citenamefont {He}, \citenamefont {Niu}, \citenamefont {Zhang}, \citenamefont
  {Ren}, \citenamefont {Liu},\ and\ \citenamefont {Jo}}]{Song19}%
  \BibitemOpen
  \bibfield  {author} {\bibinfo {author} {\bibfnamefont {Bo}~\bibnamefont
  {Song}}, \bibinfo {author} {\bibfnamefont {Chengdong}\ \bibnamefont {He}},
  \bibinfo {author} {\bibfnamefont {Sen}\ \bibnamefont {Niu}}, \bibinfo
  {author} {\bibfnamefont {Long}\ \bibnamefont {Zhang}}, \bibinfo {author}
  {\bibfnamefont {Zejian}\ \bibnamefont {Ren}}, \bibinfo {author}
  {\bibfnamefont {Xiong-Jun}\ \bibnamefont {Liu}}, \ and\ \bibinfo {author}
  {\bibfnamefont {Gyu-Boong}\ \bibnamefont {Jo}},\ }\bibfield  {title}
  {\enquote {\bibinfo {title} {Observation of nodal-line semimetal with
  ultracold fermions in an optical lattice},}\ }\href {\doibase
  10.1038/s41567-019-0564-y} {\bibfield  {journal} {\bibinfo  {journal} {Nature
  Physics}\ }\textbf {\bibinfo {volume} {15}},\ \bibinfo {pages} {911--916}
  (\bibinfo {year} {2019})}\BibitemShut {NoStop}%
\bibitem [{\citenamefont {Greco}\ and\ \citenamefont
  {Schnyder}(2018)}]{Greco18}%
  \BibitemOpen
  \bibfield  {author} {\bibinfo {author} {\bibfnamefont {Andr\'es}\
  \bibnamefont {Greco}}\ and\ \bibinfo {author} {\bibfnamefont {Andreas~P.}\
  \bibnamefont {Schnyder}},\ }\bibfield  {title} {\enquote {\bibinfo {title}
  {Mechanism for unconventional superconductivity in the hole-doped
  {R}ashba-{H}ubbard model},}\ }\href {\doibase 10.1103/PhysRevLett.120.177002}
  {\bibfield  {journal} {\bibinfo  {journal} {Phys. Rev. Lett.}\ }\textbf
  {\bibinfo {volume} {120}},\ \bibinfo {pages} {177002} (\bibinfo {year}
  {2018})}\BibitemShut {NoStop}%
\bibitem [{\citenamefont {Rashba}(1960)}]{Rashba60}%
  \BibitemOpen
  \bibfield  {author} {\bibinfo {author} {\bibfnamefont {E}~\bibnamefont
  {Rashba}},\ }\bibfield  {title} {\enquote {\bibinfo {title} {Properties of
  semiconductors with an extremum loop. 1. cyclotron and combinational
  resonance in a magnetic field perpendicular to the plane of the loop},}\
  }\href@noop {} {\bibfield  {journal} {\bibinfo  {journal} {Sov. Phys. Solid
  State}\ }\textbf {\bibinfo {volume} {2}},\ \bibinfo {pages} {1109--1122}
  (\bibinfo {year} {1960})}\BibitemShut {NoStop}%
\bibitem [{\citenamefont {Zhang}\ \emph {et~al.}(2015)\citenamefont {Zhang},
  \citenamefont {Wu}, \citenamefont {Li}, \citenamefont {Wen}, \citenamefont
  {Sun},\ and\ \citenamefont {Ji}}]{Zhang15}%
  \BibitemOpen
  \bibfield  {author} {\bibinfo {author} {\bibfnamefont {Xin}\ \bibnamefont
  {Zhang}}, \bibinfo {author} {\bibfnamefont {Wei}\ \bibnamefont {Wu}},
  \bibinfo {author} {\bibfnamefont {Gang}\ \bibnamefont {Li}}, \bibinfo
  {author} {\bibfnamefont {Lin}\ \bibnamefont {Wen}}, \bibinfo {author}
  {\bibfnamefont {Qing}\ \bibnamefont {Sun}}, \ and\ \bibinfo {author}
  {\bibfnamefont {An-Chun}\ \bibnamefont {Ji}},\ }\bibfield  {title} {\enquote
  {\bibinfo {title} {Phase diagram of interacting {F}ermi gas in spin-orbit
  coupled square lattices},}\ }\href {\doibase 10.1088/1367-2630/17/7/073036}
  {\bibfield  {journal} {\bibinfo  {journal} {New J. Phys.}\ }\textbf {\bibinfo
  {volume} {17}},\ \bibinfo {pages} {073036} (\bibinfo {year}
  {2015})}\BibitemShut {NoStop}%
\bibitem [{\citenamefont {Brosco}\ and\ \citenamefont
  {Capone}(2020)}]{Brosco20}%
  \BibitemOpen
  \bibfield  {author} {\bibinfo {author} {\bibfnamefont {Valentina}\
  \bibnamefont {Brosco}}\ and\ \bibinfo {author} {\bibfnamefont {Massimo}\
  \bibnamefont {Capone}},\ }\bibfield  {title} {\enquote {\bibinfo {title}
  {Rashba-metal to {M}ott-insulator transition},}\ }\href {\doibase
  10.1103/PhysRevB.101.235149} {\bibfield  {journal} {\bibinfo  {journal}
  {Phys. Rev. B}\ }\textbf {\bibinfo {volume} {101}},\ \bibinfo {pages}
  {235149} (\bibinfo {year} {2020})}\BibitemShut {NoStop}%
\bibitem [{\citenamefont {Mii}\ \emph {et~al.}(2014)\citenamefont {Mii},
  \citenamefont {Shima}, \citenamefont {Kano},\ and\ \citenamefont
  {Makoshi}}]{Mii14}%
  \BibitemOpen
  \bibfield  {author} {\bibinfo {author} {\bibfnamefont {Takashi}\ \bibnamefont
  {Mii}}, \bibinfo {author} {\bibfnamefont {Nobuyuki}\ \bibnamefont {Shima}},
  \bibinfo {author} {\bibfnamefont {Koichi}\ \bibnamefont {Kano}}, \ and\
  \bibinfo {author} {\bibfnamefont {Kenji}\ \bibnamefont {Makoshi}},\
  }\bibfield  {title} {\enquote {\bibinfo {title} {Spin-orbit interaction in
  the tight-binding model -- toward the comprehension of the {R}ashba effect at
  surfaces},}\ }\href {\doibase 10.7566/JPSJ.83.064706} {\bibfield  {journal}
  {\bibinfo  {journal} {Journal of the Physical Society of Japan}\ }\textbf
  {\bibinfo {volume} {83}},\ \bibinfo {pages} {064706} (\bibinfo {year}
  {2014})}\BibitemShut {NoStop}%
\bibitem [{\citenamefont {Dzierzawa}(1992)}]{Dzierzawa92}%
  \BibitemOpen
  \bibfield  {author} {\bibinfo {author} {\bibfnamefont {M}~\bibnamefont
  {Dzierzawa}},\ }\bibfield  {title} {\enquote {\bibinfo {title}
  {Hartree-{F}ock theory of spiral magnetic order in the 2-d {H}ubbard
  model},}\ }\href {\doibase https://doi.org/10.1007/BF01323546} {\bibfield
  {journal} {\bibinfo  {journal} {Z. Physik B - Condensed Matter}\ }\textbf
  {\bibinfo {volume} {86}},\ \bibinfo {pages} {49} (\bibinfo {year}
  {1992})}\BibitemShut {NoStop}%
\bibitem [{\citenamefont {Costa}\ \emph {et~al.}(2017)\citenamefont {Costa},
  \citenamefont {\surname{Pimentel de Lima}},\ and\ \citenamefont {\surname{dos
  Santos}}}]{Costa17a}%
  \BibitemOpen
  \bibfield  {author} {\bibinfo {author} {\bibfnamefont {N.~C.}\ \bibnamefont
  {Costa}}, \bibinfo {author} {\bibfnamefont {J.}~\bibnamefont
  {\surname{Pimentel de Lima}}}, \ and\ \bibinfo {author} {\bibfnamefont
  {R.~R.}\ \bibnamefont {\surname{dos Santos}}},\ }\bibfield  {title} {\enquote
  {\bibinfo {title} {Spiral magnetic phases on the {K}ondo lattice model: A
  {H}artree-{F}ock approach},}\ }\href {\doibase
  https://doi.org/10.1016/j.jmmm.2016.09.061} {\bibfield  {journal} {\bibinfo
  {journal} {J. Magn. Magn. Mater}\ }\textbf {\bibinfo {volume} {423}},\
  \bibinfo {pages} {74--83} (\bibinfo {year} {2017})}\BibitemShut {NoStop}%
\bibitem [{\citenamefont {Costa}\ \emph {et~al.}(2018)\citenamefont {Costa},
  \citenamefont {\surname{Pimentel de Lima}}, \citenamefont {Paiva},
  \citenamefont {ElMassalami},\ and\ \citenamefont {dos Santos}}]{Costa18a}%
  \BibitemOpen
  \bibfield  {author} {\bibinfo {author} {\bibfnamefont {N.~C.}\ \bibnamefont
  {Costa}}, \bibinfo {author} {\bibfnamefont {J.}~\bibnamefont
  {\surname{Pimentel de Lima}}}, \bibinfo {author} {\bibfnamefont
  {T.}~\bibnamefont {Paiva}}, \bibinfo {author} {\bibfnamefont
  {M.}~\bibnamefont {ElMassalami}}, \ and\ \bibinfo {author} {\bibfnamefont
  {R.~R.}\ \bibnamefont {dos Santos}},\ }\bibfield  {title} {\enquote {\bibinfo
  {title} {A mean-field approach to {K}ondo-attractive-{H}ubbard model},}\
  }\href {\doibase 10.1088/1361-648x/aaa1ab} {\bibfield  {journal} {\bibinfo
  {journal} {J. Phys. Condens. Matter}\ }\textbf {\bibinfo {volume} {30}},\
  \bibinfo {pages} {045602} (\bibinfo {year} {2018})}\BibitemShut {NoStop}%
\bibitem [{\citenamefont {dos Anjos Sousa-J\'unior}\ \emph
  {et~al.}(2020)\citenamefont {dos Anjos Sousa-J\'unior}, \citenamefont
  {de~Lima}, \citenamefont {Costa},\ and\ \citenamefont {dos
  Santos}}]{SOusaJr20}%
  \BibitemOpen
  \bibfield  {author} {\bibinfo {author} {\bibfnamefont {Sebasti\~ao}\
  \bibnamefont {dos Anjos Sousa-J\'unior}}, \bibinfo {author} {\bibfnamefont
  {Jos\'e~P.}\ \bibnamefont {de~Lima}}, \bibinfo {author} {\bibfnamefont
  {Natanael~C.}\ \bibnamefont {Costa}}, \ and\ \bibinfo {author} {\bibfnamefont
  {Raimundo~R.}\ \bibnamefont {dos Santos}},\ }\bibfield  {title} {\enquote
  {\bibinfo {title} {Superconducting {K}ondo phase in an orbitally separated
  bilayer},}\ }\href {\doibase 10.1103/PhysRevResearch.2.033168} {\bibfield
  {journal} {\bibinfo  {journal} {Phys. Rev. Research}\ }\textbf {\bibinfo
  {volume} {2}},\ \bibinfo {pages} {033168} (\bibinfo {year}
  {2020})}\BibitemShut {NoStop}%
\bibitem [{\citenamefont {Li}\ \emph {et~al.}(2011)\citenamefont {Li},
  \citenamefont {Covaci}, \citenamefont {Berciu}, \citenamefont {Baillie},\
  and\ \citenamefont {Marsiglio}}]{Li11}%
  \BibitemOpen
  \bibfield  {author} {\bibinfo {author} {\bibfnamefont {Zhou}\ \bibnamefont
  {Li}}, \bibinfo {author} {\bibfnamefont {L.}~\bibnamefont {Covaci}}, \bibinfo
  {author} {\bibfnamefont {M.}~\bibnamefont {Berciu}}, \bibinfo {author}
  {\bibfnamefont {D.}~\bibnamefont {Baillie}}, \ and\ \bibinfo {author}
  {\bibfnamefont {F.}~\bibnamefont {Marsiglio}},\ }\bibfield  {title} {\enquote
  {\bibinfo {title} {Impact of spin-orbit coupling on the holstein polaron},}\
  }\href {\doibase 10.1103/PhysRevB.83.195104} {\bibfield  {journal} {\bibinfo
  {journal} {Phys. Rev. B}\ }\textbf {\bibinfo {volume} {83}},\ \bibinfo
  {pages} {195104} (\bibinfo {year} {2011})}\BibitemShut {NoStop}%
\bibitem [{\citenamefont {Ptok}\ \emph {et~al.}(2018)\citenamefont {Ptok},
  \citenamefont {Rodr\'{\i}guez},\ and\ \citenamefont {Kapcia}}]{Ptok18}%
  \BibitemOpen
  \bibfield  {author} {\bibinfo {author} {\bibfnamefont {Andrzej}\ \bibnamefont
  {Ptok}}, \bibinfo {author} {\bibfnamefont {Karen}\ \bibnamefont
  {Rodr\'{\i}guez}}, \ and\ \bibinfo {author} {\bibfnamefont {Konrad~Jerzy}\
  \bibnamefont {Kapcia}},\ }\bibfield  {title} {\enquote {\bibinfo {title}
  {Superconducting monolayer deposited on substrate: Effects of the spin-orbit
  coupling induced by proximity effects},}\ }\href {\doibase
  10.1103/PhysRevMaterials.2.024801} {\bibfield  {journal} {\bibinfo  {journal}
  {Phys. Rev. Materials}\ }\textbf {\bibinfo {volume} {2}},\ \bibinfo {pages}
  {024801} (\bibinfo {year} {2018})}\BibitemShut {NoStop}%
\bibitem [{\citenamefont {Hutchinson}\ \emph {et~al.}(2018)\citenamefont
  {Hutchinson}, \citenamefont {Hirsch},\ and\ \citenamefont
  {Marsiglio}}]{Hutchinson18}%
  \BibitemOpen
  \bibfield  {author} {\bibinfo {author} {\bibfnamefont {Joel}\ \bibnamefont
  {Hutchinson}}, \bibinfo {author} {\bibfnamefont {J.~E.}\ \bibnamefont
  {Hirsch}}, \ and\ \bibinfo {author} {\bibfnamefont {Frank}\ \bibnamefont
  {Marsiglio}},\ }\bibfield  {title} {\enquote {\bibinfo {title} {Enhancement
  of superconducting ${T}_{c}$ due to the spin-orbit interaction},}\ }\href
  {\doibase 10.1103/PhysRevB.97.184513} {\bibfield  {journal} {\bibinfo
  {journal} {Phys. Rev. B}\ }\textbf {\bibinfo {volume} {97}},\ \bibinfo
  {pages} {184513} (\bibinfo {year} {2018})}\BibitemShut {NoStop}%
\bibitem [{\citenamefont {Igoshev}\ \emph {et~al.}(2015)\citenamefont
  {Igoshev}, \citenamefont {Timirgazin}, \citenamefont {Gilmutdinov},
  \citenamefont {Arzhnikov},\ and\ \citenamefont {Irkhin}}]{Igoshev15}%
  \BibitemOpen
  \bibfield  {author} {\bibinfo {author} {\bibfnamefont {PA}~\bibnamefont
  {Igoshev}}, \bibinfo {author} {\bibfnamefont {MA}~\bibnamefont {Timirgazin}},
  \bibinfo {author} {\bibfnamefont {VF}~\bibnamefont {Gilmutdinov}}, \bibinfo
  {author} {\bibfnamefont {AK}~\bibnamefont {Arzhnikov}}, \ and\ \bibinfo
  {author} {\bibfnamefont {V~Yu}\ \bibnamefont {Irkhin}},\ }\bibfield  {title}
  {\enquote {\bibinfo {title} {Spiral magnetism in the single-band hubbard
  model: the hartre--fock and slave-boson approaches},}\ }\href@noop {}
  {\bibfield  {journal} {\bibinfo  {journal} {Journal of Physics: Condensed
  Matter}\ }\textbf {\bibinfo {volume} {27}},\ \bibinfo {pages} {446002}
  (\bibinfo {year} {2015})}\BibitemShut {NoStop}%
\bibitem [{\citenamefont {Hirsch}(1985)}]{Hirsch85}%
  \BibitemOpen
  \bibfield  {author} {\bibinfo {author} {\bibfnamefont {J.~E.}\ \bibnamefont
  {Hirsch}},\ }\bibfield  {title} {\enquote {\bibinfo {title} {Two-dimensional
  {H}ubbard model: Numerical simulation study},}\ }\href {\doibase
  10.1103/PhysRevB.31.4403} {\bibfield  {journal} {\bibinfo  {journal} {Phys.
  Rev. B}\ }\textbf {\bibinfo {volume} {31}},\ \bibinfo {pages} {4403--4419}
  (\bibinfo {year} {1985})}\BibitemShut {NoStop}%
\bibitem [{\citenamefont {Hu}\ \emph {et~al.}(2017)\citenamefont {Hu},
  \citenamefont {Scalettar}, \citenamefont {Huang},\ and\ \citenamefont
  {Moritz}}]{Hu17}%
  \BibitemOpen
  \bibfield  {author} {\bibinfo {author} {\bibfnamefont {Wenjian}\ \bibnamefont
  {Hu}}, \bibinfo {author} {\bibfnamefont {Richard~T.}\ \bibnamefont
  {Scalettar}}, \bibinfo {author} {\bibfnamefont {Edwin~W.}\ \bibnamefont
  {Huang}}, \ and\ \bibinfo {author} {\bibfnamefont {Brian}\ \bibnamefont
  {Moritz}},\ }\bibfield  {title} {\enquote {\bibinfo {title} {Effects of an
  additional conduction band on the singlet-antiferromagnet competition in the
  periodic anderson model},}\ }\href {\doibase 10.1103/PhysRevB.95.235122}
  {\bibfield  {journal} {\bibinfo  {journal} {Phys. Rev. B}\ }\textbf {\bibinfo
  {volume} {95}},\ \bibinfo {pages} {235122} (\bibinfo {year}
  {2017})}\BibitemShut {NoStop}%
\bibitem [{\citenamefont {Penn}(1966)}]{Penn66}%
  \BibitemOpen
  \bibfield  {author} {\bibinfo {author} {\bibfnamefont {David~R.}\
  \bibnamefont {Penn}},\ }\bibfield  {title} {\enquote {\bibinfo {title}
  {Stability theory of the magnetic phases for a simple model of the transition
  metals},}\ }\href {\doibase 10.1103/PhysRev.142.350} {\bibfield  {journal}
  {\bibinfo  {journal} {Phys. Rev.}\ }\textbf {\bibinfo {volume} {142}},\
  \bibinfo {pages} {350--365} (\bibinfo {year} {1966})}\BibitemShut {NoStop}%
\bibitem [{\citenamefont {Bertussi}\ \emph {et~al.}(2009)\citenamefont
  {Bertussi}, \citenamefont {Malvezzi}, \citenamefont {Paiva},\ and\
  \citenamefont {dos Santos}}]{Bertussi09}%
  \BibitemOpen
  \bibfield  {author} {\bibinfo {author} {\bibfnamefont {Pedro~R.}\
  \bibnamefont {Bertussi}}, \bibinfo {author} {\bibfnamefont {Andr\'e~L.}\
  \bibnamefont {Malvezzi}}, \bibinfo {author} {\bibfnamefont {Thereza}\
  \bibnamefont {Paiva}}, \ and\ \bibinfo {author} {\bibfnamefont {Raimundo~R.}\
  \bibnamefont {dos Santos}},\ }\bibfield  {title} {\enquote {\bibinfo {title}
  {Kondo--attractive-{H}ubbard model for the ordering of local magnetic moments
  in superconductors},}\ }\href {\doibase 10.1103/PhysRevB.79.220513}
  {\bibfield  {journal} {\bibinfo  {journal} {Phys. Rev. B}\ }\textbf {\bibinfo
  {volume} {79}},\ \bibinfo {pages} {220513} (\bibinfo {year}
  {2009})}\BibitemShut {NoStop}%
\bibitem [{\citenamefont {ElMassalami}\ \emph {et~al.}(2012)\citenamefont
  {ElMassalami}, \citenamefont {Takeya}, \citenamefont {Ouladdiaf},
  \citenamefont {Maia~Filho}, \citenamefont {Gomes}, \citenamefont {Paiva},\
  and\ \citenamefont {dos Santos}}]{ElMassalami12}%
  \BibitemOpen
  \bibfield  {author} {\bibinfo {author} {\bibfnamefont {M.}~\bibnamefont
  {ElMassalami}}, \bibinfo {author} {\bibfnamefont {H.}~\bibnamefont {Takeya}},
  \bibinfo {author} {\bibfnamefont {B.}~\bibnamefont {Ouladdiaf}}, \bibinfo
  {author} {\bibfnamefont {R.}~\bibnamefont {Maia~Filho}}, \bibinfo {author}
  {\bibfnamefont {A.~M.}\ \bibnamefont {Gomes}}, \bibinfo {author}
  {\bibfnamefont {T.}~\bibnamefont {Paiva}}, \ and\ \bibinfo {author}
  {\bibfnamefont {R.~R.}\ \bibnamefont {dos Santos}},\ }\bibfield  {title}
  {\enquote {\bibinfo {title} {Tuning in magnetic modes in
  {T}b({C}o$_{x}${N}i$_{1-x}$)$_{2}${B}$_2${C}: From longitudinal spin-density
  waves to simple ferromagnetism},}\ }\href {\doibase
  10.1103/PhysRevB.85.174412} {\bibfield  {journal} {\bibinfo  {journal} {Phys.
  Rev. B}\ }\textbf {\bibinfo {volume} {85}},\ \bibinfo {pages} {174412}
  (\bibinfo {year} {2012})}\BibitemShut {NoStop}%
\bibitem [{\citenamefont {ElMassalami}\ \emph {et~al.}(2013)\citenamefont
  {ElMassalami}, \citenamefont {Gomes}, \citenamefont {Paiva}, \citenamefont
  {{dos Santos}},\ and\ \citenamefont {Takeya}}]{ElMassalami13}%
  \BibitemOpen
  \bibfield  {author} {\bibinfo {author} {\bibfnamefont {M.}~\bibnamefont
  {ElMassalami}}, \bibinfo {author} {\bibfnamefont {A.M.}\ \bibnamefont
  {Gomes}}, \bibinfo {author} {\bibfnamefont {T.}~\bibnamefont {Paiva}},
  \bibinfo {author} {\bibfnamefont {R.R.}\ \bibnamefont {{dos Santos}}}, \ and\
  \bibinfo {author} {\bibfnamefont {H.}~\bibnamefont {Takeya}},\ }\bibfield
  {title} {\enquote {\bibinfo {title} {Evolution of magnetism in
  {T}b({C}o$_x${N}i$_{1-x}$)$_2${B}$_2${C}},}\ }\href {\doibase
  https://doi.org/10.1016/j.jmmm.2013.01.044} {\bibfield  {journal} {\bibinfo
  {journal} {Journal of Magnetism and Magnetic Materials}\ }\textbf {\bibinfo
  {volume} {335}},\ \bibinfo {pages} {163--171} (\bibinfo {year}
  {2013})}\BibitemShut {NoStop}%
\bibitem [{\citenamefont {ElMassalami}\ \emph {et~al.}(2014)\citenamefont
  {ElMassalami}, \citenamefont {Takeya}, \citenamefont {Ouladdiaf},
  \citenamefont {Gomes}, \citenamefont {Paiva},\ and\ \citenamefont {{dos
  Santos}}}]{ElMassalami14}%
  \BibitemOpen
  \bibfield  {author} {\bibinfo {author} {\bibfnamefont {M.}~\bibnamefont
  {ElMassalami}}, \bibinfo {author} {\bibfnamefont {H.}~\bibnamefont {Takeya}},
  \bibinfo {author} {\bibfnamefont {B.}~\bibnamefont {Ouladdiaf}}, \bibinfo
  {author} {\bibfnamefont {A.M.}\ \bibnamefont {Gomes}}, \bibinfo {author}
  {\bibfnamefont {T.}~\bibnamefont {Paiva}}, \ and\ \bibinfo {author}
  {\bibfnamefont {R.R.}\ \bibnamefont {{dos Santos}}},\ }\bibfield  {title}
  {\enquote {\bibinfo {title} {Evolution of magnetic layers stacking sequence
  within the magnetic structure of {H}o({C}o$_x${N}i$_{1-x}$)$_2${B}$_2${C}},}\
  }\href {\doibase https://doi.org/10.1016/j.jmmm.2014.07.026} {\bibfield
  {journal} {\bibinfo  {journal} {Journal of Magnetism and Magnetic Materials}\
  }\textbf {\bibinfo {volume} {372}},\ \bibinfo {pages} {74--78} (\bibinfo
  {year} {2014})}\BibitemShut {NoStop}%
\bibitem [{\citenamefont {Liu}(1994)}]{Liu94}%
  \BibitemOpen
  \bibfield  {author} {\bibinfo {author} {\bibfnamefont {Bang-Gui}\
  \bibnamefont {Liu}},\ }\bibfield  {title} {\enquote {\bibinfo {title}
  {Incommensurate antiferromagnetic orders and peculiar electronic structures
  in the doped {H}ubbard model},}\ }\href {\doibase 10.1088/0953-8984/6/30/001}
  {\bibfield  {journal} {\bibinfo  {journal} {Journal of Physics: Condensed
  Matter}\ }\textbf {\bibinfo {volume} {6}},\ \bibinfo {pages} {L415--L421}
  (\bibinfo {year} {1994})}\BibitemShut {NoStop}%
\bibitem [{Note1()}]{Note1}%
  \BibitemOpen
  \bibinfo {note} {One should keep in mind that the exchange symmetry $q_x
  \leftrightarrow q_y$ applies, \protect \textit {i.e.}\ the solutions
  $\protect \textbf {q}=(q_x,q_y)$ and $\protect \textbf {q}=(q_y,q_x)$ are
  degenerate.}\BibitemShut {Stop}%
\bibitem [{\citenamefont {Park}\ \emph {et~al.}(2013)\citenamefont {Park},
  \citenamefont {Kim}, \citenamefont {Lee},\ and\ \citenamefont
  {Han}}]{Park2013}%
  \BibitemOpen
  \bibfield  {author} {\bibinfo {author} {\bibfnamefont {Jin-Hong}\
  \bibnamefont {Park}}, \bibinfo {author} {\bibfnamefont {Choong~H.}\
  \bibnamefont {Kim}}, \bibinfo {author} {\bibfnamefont {Hyun-Woo}\
  \bibnamefont {Lee}}, \ and\ \bibinfo {author} {\bibfnamefont {Jung~Hoon}\
  \bibnamefont {Han}},\ }\bibfield  {title} {\enquote {\bibinfo {title}
  {Orbital chirality and rashba interaction in magnetic bands},}\ }\href
  {\doibase 10.1103/PhysRevB.87.041301} {\bibfield  {journal} {\bibinfo
  {journal} {Phys. Rev. B}\ }\textbf {\bibinfo {volume} {87}},\ \bibinfo
  {pages} {041301} (\bibinfo {year} {2013})}\BibitemShut {NoStop}%
\end{thebibliography}%

\end{document}